\documentclass[12pt,a4paper]{article}
\usepackage[dvips,colorlinks,bookmarksopen,bookmarksnumbered,linkcolor=black,citecolor=black,urlcolor=black,breaklinks]{hyperref}
\usepackage{graphicx}
\usepackage{breakurl}
\usepackage[round]{natbib}
\newcommand*{\doi}[1]{\href{http://dx.doi.org/#1}{doi:#1}}

\begin{document}

\title{Meta-analysis of few small studies\\ in orphan diseases}

\author{T.~Friede\thanks{\textit{correspondence to}: Tim Friede, Department of Medical Statistics, University Medical Center G\"{o}ttingen, Germany; email: \texttt{tim.friede@med.uni-goettingen.de}}, C.~R\"{o}ver, S.~Wandel, B.~Neuenschwander}

\date{September 30, 2015}

\maketitle

\begin{abstract}
Meta-analyses in orphan diseases and small populations generally face particular problems including small numbers of studies, small study sizes, and heterogeneity of results. However, the heterogeneity is difficult to estimate if only very few studies are included. Motivated by a systematic review in immunosuppression following liver transplantation in children we investigate the properties of a range of commonly used frequentist and Bayesian procedures in extensive simulation studies. Furthermore, the consequences for interval estimation of the common treatment effect in random effects meta-analysis are assessed. The Bayesian credibility intervals using weakly informative priors for the between-trial heterogeneity exhibited coverage probabilities in excess of the nominal level for a range of scenarios considered. However, they tended to be shorter than those obtained by the Knapp-Hartung method, which were also conservative. In contrast, methods based on normal quantiles exhibited coverages well below the nominal levels in many scenarios. With very few studies, the performance of the Bayesian credibility intervals is of course sensitive to the specification of the prior for the between trial heterogeneity. In conclusion, the use of weakly informative priors as exemplified by half-normal priors (with scale 0.5 or 1.0) for log odds ratios is recommended for applications in rare diseases.
\end{abstract}

\section{Introduction} \label{sec:intro}

In the European Union a disease affecting 5 in 10,000 people is considered rare, whereas in the USA a condition affecting fewer than 200,000 people is defined as rare. As these examples show, no universal definition of rare diseases (also referred to as orphan disease) exists \citep{Aronson2006}. It is estimated that 6,000 to 8,000 rare diseases are known today with numbers increasing as more diseases are discovered. Many of these have a genetic component, are chronic as well as life-threatening and affect children (see e.g.\ \href{http://rarediseases.org}{\url{rarediseases.org}} or \href{http://www.orpha.net}{\url{www.orpha.net}}). Examples of rare diseases include childhood cancers, amyotrophic lateral sclerosis (ALS), and Creutzfeldt-Jacob disease (CJD). More generally, small populations can occur by rare conditions or by stratification of more common diseases by e.g.\ genetic markers.  

Since rare diseases and small populations pose particular problems to design, conduct and analysis of clinical research due to the small sizes, various efforts have been undertaken in tailoring methods specific for the application in these populations \citep{GagneEtAl2014}. Recent publications such as \citet{HampsonEtAl2014,SpeiserEtAl2015} demonstrate that this is an ongoing effort which is also reflected in a recent funding initiative of the European Commission supporting three research networks in developing clinical research methodology suitable for rare diseases and small populations, namely ``Integrated Design and AnaLysis of small population group trials'' (IDEAL), ``Advances in Small Trials dEsign for Regulatory Innovation and eXcellence'' (ASTERIX) and ``Innovative methodology for small population research'' (InSPiRe). Also, regulatory authorities acknowledge the need for innovative approaches to clinical trials in rare and very rare diseases \citep{EMEA2006}. To our knowledge, however, neither have the standard methods of meta-analysis been assessed for their suitability to be applied in small populations and rare diseases nor have any specific methods for meta-analysis been developed for these populations. This is surprising, since it is generally accepted that meta-analytic methods are a powerful tool to guide objective decision-making by allowing for the formal, statistical combination of information to merge data from individual experiments to a joint result.

There are some specific challenges for meta-analyses in small populations and rare diseases. As will be seen in the motivating example of Section~\ref{sec:example}, the number of studies included in meta-analyses is typically small, and the studies themselves are rather small with many of them being inconclusive. In this situation a formal synthesis of the available evidence is highly desirable. However, the study designs often vary, for instance with regard to the type of control groups (historical vs. concurrent controls) and treatment allocation (non-randomized vs. randomized; allocation ratio), making between-trial heterogeneity with regard to the treatment effects very likely. In fact, smaller studies have empirically been found to exhibit more heterogeneity than large ones \citep{InthoutEtAl2015}.  The situation gets even more difficult for rare events \citep{SweetingSuttonLambert2004,BradburnEtAl2007,Kuss2015}, but this is not our scope here.

Many meta-analyses may be approached using approximately normal effect estimates and a normal random-effects model, the normal-normal hierarchical model \citep{HedgesOlkin}. The between-study heterogeneity plays a central role in this context.  Estimation of the heterogeneity variance component based on only a few studies however is particularly challenging, as the resulting uncertainty is often substantial and consistent with small to large heterogeneity. Proper accounting for this uncertainty is crucial when estimating effect sizes.  Especially when the analysis is based on few studies, which is a common problem not only for orphan diseases, the utilization of a-priori information on heterogeneity may be helpful \citep{SuttonAbrams2001}.

Our aim is to assess the properties of Bayesian and popular frequentist methods in a simulation study covering typical scenarios for rare disease as well as in a case study.  We have developed a method to substantially ease the computational burden of a Bayesian random-effects meta-analysis, which avoids Markov chain Monte Carlo computations and facilitates large-scale simulations. The method has been implemented in the \texttt{bayesmeta} \textsf{R}~package to be released soon.

The paper is organized as follows. In Section~\ref{sec:example} we introduce a case study in paediatric transplantation motivating our investigations. In Section~\ref{sec:meth} the statistical model as well as the methods for meta-analysis are introduced that are compared in an extensive simulation study in Section~\ref{sec:sims}. The findings are illustrated by revisiting the systematic review in paediatric transplantation in Section~\ref{sec:example_rev}. We close with a brief discussion.

\section{A case study in paediatric transplantation} \label{sec:example}

Several rare paediatric liver diseases can nowadays be successfully treated by liver transplantation with good long-term outcomes \citep{SpadaEtAl2009,Kosolo2013}. One key component of a successful treatment is effective and safe immunosuppresion after transplantation. A number of immunosuppressant drugs with different mechanisms of action and associated adverse event profiles are available and are given either as monotherapy or in combination (for an overview see e.g.\ Table~1 in \citet{Kosolo2013}). Relatively new are the Interleukin-2 receptor antibodies (IL-2RA) Basiliximab and Daclizumab. Recently, \citet{CrinsEtAl2014} conducted a systematic review of controlled, but not necessarily randomized studies of IL-2RA in paediatric liver transplantation.  Primary outcomes were acute rejections, steroid-resistant rejections, graft-loss and death. For illustrative purposes we focus here on acute and steroid-resistant rejections. A total of six studies were included in a meta-analysis assessing the risk of acute rejections on treatments with IL-2RA in comparison with control, whereas only three of these six studies reported data on steroid-resistant rejections.

\begin{figure}[ht]
\begin{center}
\includegraphics[width=10cm]{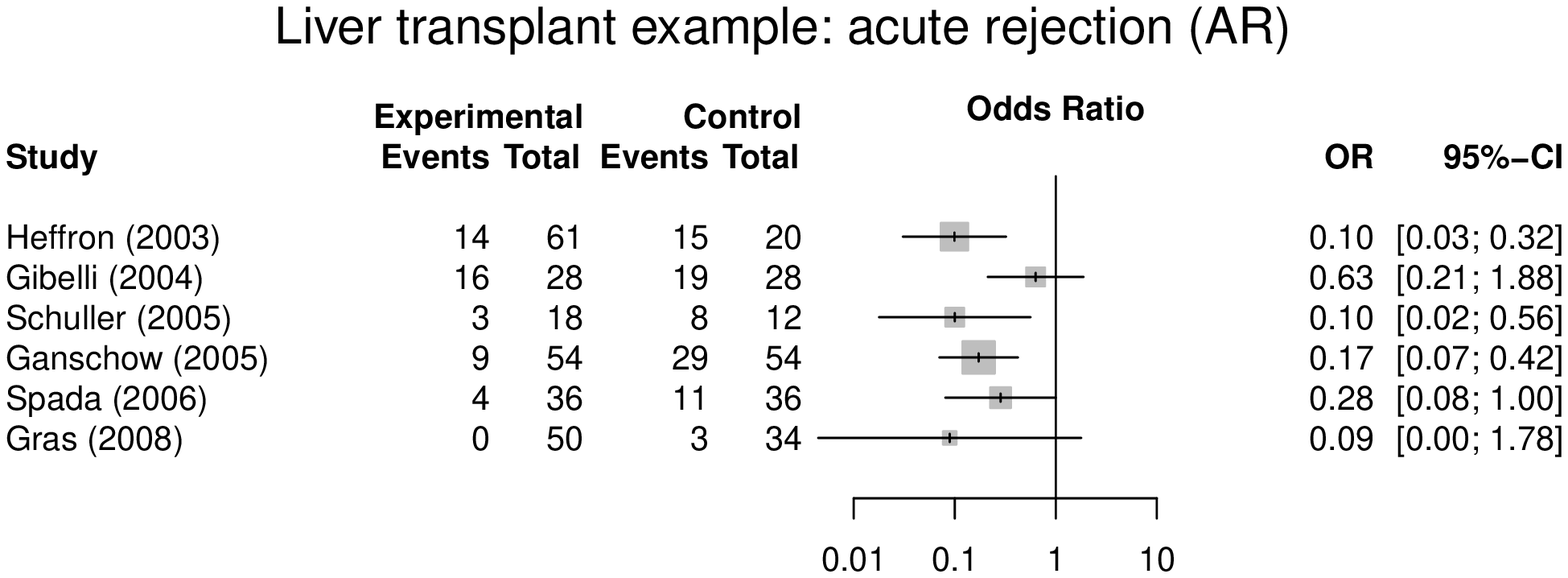} \\[2ex]
\includegraphics[width=10cm]{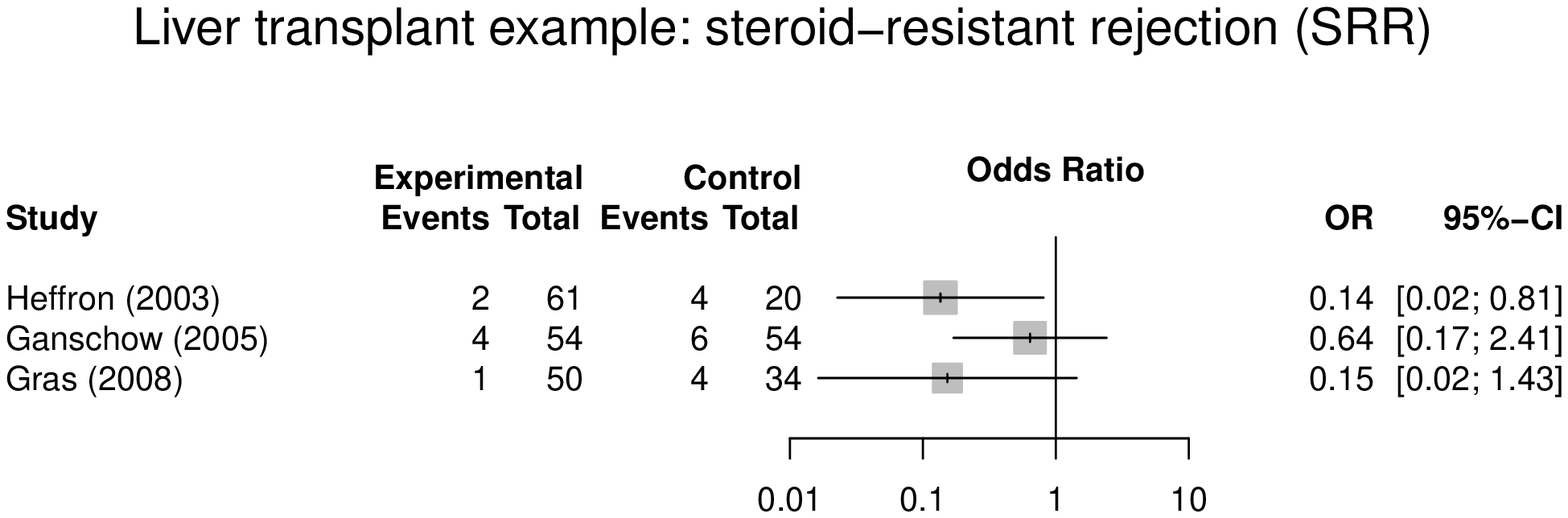}
\caption{Forest plots of odds ratios for acute rejections and steroid-resistant rejections based on data from the systematic review by \citet{CrinsEtAl2014}.} \label{fig:crins}
\end{center}
\end{figure}

Crins et al.\ used the DerSimonian-Laird method and the restricted max\-i\-mum-likelihood (\textsl{REML}) approach to estimate the between-study variance. Although they reported treatment effects in terms of relative risks we use here the odds ratio as it is the more commonly used measure. Figure~\ref{fig:crins} gives the numbers of patients and events by treatment group, and depicts the odds ratios comparing IL-2RA treament with control in forest plots. The 95\% confidence intervals of the odds ratios were computed using normal approximation on the log-scale (zero cell entries here were treated by adding $\frac{1}{2}$ to each cell of the corresponding contingency table).

The specific problems with meta-analyses in rare diseases outlined in Section~\ref{sec:intro} are prominent here. First, the number of controlled studies is with 6 and 3 rather small, although the search was not restricted to randomized studies only. Second, the total sample sizes varied from only 30 to 108. Third, there are some marked differences in the designs employed. For instance, only the studies by Heffron (2003) and Spada (2006) were randomized, and some of the non-randomized studies used concurrent controls whereas other relied on historical control data. Finally, the allocation ratios ranged between 1:1 to 3:1. These variations in design lead to some heterogeneity in the results as is apparent from the forest plots. In the next section we will summarize a number of methods to quantify this heterogeneity. 

\section{Methods} \label{sec:meth}
The vast majority of meta-analyses use common effect (fixed effect) or random effects models. For the latter, the normal-normal hierarchical model \textsl{(NNHM)} is the most popular. It has two parts: a sampling and a parameter model. The sampling model assumes approximately normally distributed estimates $Y_1,\ldots,Y_k$ for the trial-specific parameters $\theta_1,\ldots,\theta_k$
\begin{equation}
Y_j \vert \theta_j \sim N(\theta_j,s_j^2), \quad j=1,\ldots,k .
\label{NNHM:est}
\end{equation}
Any estimation uncertainty in the standard errors~$s_j$ is neglected here. The parameter model softens the strong (common effect) assumption of parameter equality to parameter similarity. The simplest similarity model assumes the parameters as random (or exchangeable) effects
\begin{equation}
\label{NNHM:par}
\theta_j \vert \mu, \tau \sim N(\mu,\tau^2),  \quad j=1,\ldots,k .
\end{equation}
Equivalently, this model introduces a variance component for between-trial variability, $\theta_j=\mu+\epsilon_j$, with $\epsilon_j \sim N(0,\tau^2)$.  The between-trial standard deviation $\tau$ determines the degree of similarity across parameters; the special case $\tau=0$ corresponds to the common effect model.

The scope of the \textsl{NNHM} is broad. The focus can be on the trial-specific parameters $\theta_j$, the predicted parameter $\theta_{k+1}$ in an new trial, or the mean parameter $\mu$. Here, we will be concerned with the latter. For this case, inference can be simplified by considering only the marginal model
\begin{equation}
\label{MargModel}
Y_j \vert \mu,\tau \sim N(\mu,s_j^2+\tau^2), \quad j=1,\ldots,k .
\end{equation}
The two ways to infer the parameter of interest $\mu$ and the nuisance parameter $\tau$ are classical and Bayesian. If $\tau$ were known, they would lead to the same conclusions for $\mu$. Classically,  
\begin{equation}
\label{est.mu}
 \hat{\mu} = \sum_{j=1}^k w_jY_j / \sum_{j=1}^k w_j \sim N(\mu,1/w_{+})
\end{equation}
where $w_j$ are the inverse-variance (precision) weights
\begin{equation}
\label{weights}
w_j = 1 /(s_j^2+\tau^2), \quad j=1,\ldots k,
\end{equation}
and the total precision of the estimate is the sum of the marginal precisions, $w_+ = \sum_{j=1}^k w_j$. For known $\tau$ and a non-informative (improper) prior on $\mu$, the Bayesian result is 
\begin{equation}
\label{mu:posterior}
\mu \vert Y_1,\ldots,Y_k \sim N(\sum_{j=1}^k w_jY_j / \sum_{j=1}^k w_j,1/w_{+}) .
\end{equation}
While this ``equivalence'' of classical and Bayesian results for $\mu$ is comforting, it breaks down if $\tau$ is not known. For this case, classical methods to infer $\mu$ involve two steps: 
\begin{enumerate}
\item[(1)] an estimate $\hat{\tau}$ is derived, which is then used to
  obtain the weights (\ref{weights}) and the corresponding estimate
  for $\mu$ in (\ref{est.mu});
\item[(2)]
a confidence interval is derived. In this step the uncertainty for the estimate $\hat{\tau}$ is sometimes ignored, i.e., the normal approximation in (\ref{est.mu}) with the plug-in estimate $\hat{\tau}$ is used. Or, the uncertainty of $\hat{\tau}$ is taken into account, for example by the $t_{k-1}$
approximation \citep{FollmannProschan1999}, the profile likelihood method \citep{HardyThompson1996}, or the method by Hartung and Knapp \citep{HartungKnapp2001a,HartungKnapp2001b,KnappHartung2003}. 
\end{enumerate}
For step (1) many methods have been proposed (e.g.\ \citet{DerSimonianKacker2007,ChungEtAl2013a}). In the following sections we will use four estimates:
\begin{itemize}
\item \textsl{DL}: the frequently used \textsl{DerSimonian-Laird} estimate \citep{DerSimonianLaird1986} is a moment-based estimate for $\tau$. The \textsl{DL} estimate does not require distributional assumptions and is available in closed form. However, it tends to underestimate $\tau$. It often leads to zero estimates for $\tau$ (in particular for small numbers of trials), and thus results in a common effect meta-analysis with potentially too optimistic confidence intervals for $\mu$.
\item \textsl{REML}: the \textsl{restricted (or residual) maximum likelihood estimate} has been proposed as an improvement over the standard maximum likelihood \textsl{(ML)} estimate \citep{Viechtbauer2005}. The \textsl{REML} estimate requires iterative computations. It has the advantage of being less downward biased compared to the \textsl{DL} and \textsl{ML} estimates.
\item \textsl{MP}: the \textsl{Mandel-Paule} \citep{PauleMandel1982} estimate provides an approximation to the \textsl{REML} estimate \citep{RukhinBiggerstaffVangel2000,DerSimonianKacker2007}, like the \textsl{DL} estimate it does not require distributional assumptions, and it has been recommended for use in meta-analysis \citep{VeronikiEtAl2015}.
\item \textsl{BM}: the \textsl{Bayes-modal} estimate \citep{ChungEtAl2013a} is an example of a hybrid approach. In the first step, it derives a non-zero Bayesian estimate for $\tau$ using a gamma prior with shape parameter $\alpha>1$, and then proceeds in a classical way to infer $\mu$ in step (2). Effectively, this is a penalized likelihood approach.
\end{itemize}
For step (2) we will use two methods to derive confidence intervals for $\mu$ for each of the four $\tau$ estimates. 
\begin{itemize} 
\item The simple normal approximation (\ref{est.mu}), which ignores the uncertainty for $\hat{\tau}$. The $1-\alpha$ confidence interval is constructed as $\hat \mu \pm z_{1-\alpha/2}\ \sigma_\mu$ where $z_\gamma$ is the $\gamma$-quantile of the standard normal distribution and $\sigma_\mu$ the standard error of $\hat \mu$ given by $\sigma_\mu = \sqrt{\frac{1}{w_+}}$.
\item The method by Hartung and Knapp \citep{HartungKnapp2001a,HartungKnapp2001b,KnappHartung2003}. Their method assumes an estimate $\hat{\mu}$ has been obtained (e.g.\ based on the \textsl{DL} estimate for $\tau$, which also implies weights $w_j$ in (\ref{weights})). Based on these quantities, the variance estimate for $\hat{\mu}$ is obtained as $q \sigma^2_\mu$ with
\begin{equation}
q = \frac{1}{k-1} \sum_j w_j(y_j-\hat{\mu})^2 .
\end{equation}
The variance estimator $q \sigma^2_\mu$ can be interpreted as a weighted extension of the usual empirical variance of the study-specific estimator. However, there are additional results which make $q \sigma^2_u$ attractive \citep{HartungKnapp2001a,HartungKnapp2001b,KnappHartung2003}: $q \sigma^2_\mu$ is an unbiased estimate of $\mbox{Var}(\hat{\mu})$, and $w_+(k-1)q \sigma^2_\mu$ follows a $\chi^2_{k-1}$ distribution. This, in turn, leads to the construction of an $1-\alpha$ confidence interval for $\mu$ as $\hat{\mu} \pm t_{k-1;1-\alpha/2} \ \sqrt{q}\ \sigma_\mu$. Here we use the modified version replacing $q$ by $q^\star$ with $q^\star := \max\{1, q \}$ \citep{RoeverKnappFriede2015}. Furthermore, this interval provides coverage for $\mu$ close to the nominal level \citep{SidikJonkman2003}, at least when the number of studies is sufficiently large.
\end{itemize}

In the Bayesian approach, uncertainty about $\tau$ is automatically accounted for when estimating $\mu$. However, the choice of prior matters, in particular if the number of studies is small. This case has been discussed in \citet{TurnerEtAl2015}, Section~2 of \citet{DiasEtAl2012}, and Section~6.2 of \citet{DiasEtAl2014}. While we suggest to use informative priors for $\tau$ if solid information about between-trial heterogeneity is available, in what follows we will only consider the case where prior information is weak. For this case, we recommend to use priors that put most of their probability mass to values that represent small to large between-trial heterogeneity and leave the remaining probability (e.g.\ 5\%) to values that reflect very large heterogeneity. These weakly-informative priors can be represented by half-normal, half-Cauchy, or half-t distributions \citep{SpiegelhalterEtAl,Gelman2006,PolsonScott2012}.

What constitues small to large heterogeneity depends on the parameter scale. The situation we consider here is the two-group (test vs. control) binomial case with parameters $\pi_{Tj}$ and $\pi_{Cj}$ for study $j$. The numbers of patients in the test and control group are denoted by $n_{Tj}$ and $n_{Cj}$, respectively, while the numbers of events are denoted by $r_{Tj}$ and $r_{Cj}$. Here the log-odds-ratio $\theta_j$ is used for comparison.  The corresponding approximate \textsl{NNHM} assumes exchangeable log-odds-ratios across trials
\begin{equation}
\theta_j = \log \biggl(\frac{\pi_{Tj}(1-\pi_{Cj})}{\pi_{Cj}(1-\pi_{Tj})}\biggr) \sim N(\mu,\tau^2) 
\end{equation}
and approximately normal log-odds-ratio estimates $Y_j$ with standard errors $s_j$ 
\begin{equation}
Y_j = \log \biggl(\frac{r_{Tj}(n_{Cj}-r_{Cj})}{r_{Cj}(n_{Tj}-r_{Tj})}\biggr)\mbox{,} \quad
s_j^2 = 1/r_{Tj} + 1/(n_{Tj}-r_{Tj}) + 1/r_{Cj} + 1/(n_{Cj}-r_{Cj}) \mbox{.}
\end{equation}

We will use half-normal priors $HN(\varphi)$ that cover the range of typical $\tau$ values representing small to very large heterogeneity: 0.125 (small), 0.25 (moderate), 0.5 (substantial), 1 (large), and 2 (very large); see \citet{SpiegelhalterEtAl}. A half-normal distribution $HN(\varphi)$ with scale $\varphi$ results from taking absolute values of observations sampled from a normal distribution with expectation 0 and variance $\varphi^2$. To illustrate between-trial heterogeneity for these $\tau$ values, Table~\ref{tab:tau} shows 95\%-intervals for across-trial odds-ratios (conditional on $\mu=0$). In Table~\ref{tab:tauPriors} the three ``weakly-informative'' priors are shown which we will use in the following sections. These are two half-normal priors with scale 0.5 and 1, and a uniform (0,4) prior. The latter, which puts 50\% prior probability on very large between-trial heterogeneity (in most cases a rather unrealistic assumption) will be used to assess prior sensitivity.   

\begin{table} [ht]
\begin{center}
  \caption{Between-trial heterogeneity for log-odds-ratios:
    $\tau$~values representing small to very large heterogeneity, with
    95\%-intervals for across-trial odds-ratios ($\exp(\theta_j)$).}
  \label{tab:tau}
  \begin{tabular}{r@{: }l@{\hspace{3ex}}c}
    \hline \hline
    \multicolumn{2}{c}{heterogeneity} & 95\% interval \\ \hline
          small &  $\tau=0.125$ & $0.783-\phantom{0}1.28$ \\
       moderate &  $\tau=0.25$  & $0.613-\phantom{0}1.63$ \\
    substantial &  $\tau=0.5$   & $0.325-\phantom{0}2.66$ \\
          large &  $\tau=1$     & $0.141-\phantom{0}7.10$ \\
     very large &  $\tau=2$     & $0.020-50.4\phantom{0}$ \\
    \hline \hline
  \end{tabular}
\end{center}
\end{table}

\begin{table} [ht]
\begin{center}
  \caption{Between-trial heterogeneity for log-odds-ratios: three
    priors covering small to large heterogeneity.}
  \label{tab:tauPriors}
  \begin{tabular}{lcc}
    \hline \hline
    prior distribution     & median & 95\% interval \\ \hline
    half-Normal(scale=0.5) & 0.337  & (0.016, 1.12) \\
    half-Normal(scale=1.0) & 0.674  & (0.031, 2.24) \\
    Uniform(0,4)           & 2.0    & (0.1, 3.9)    \\ \hline \hline
\end{tabular}
\end{center}
\end{table}

\section{Simulation study} \label{sec:sims}

In order to compare the performance of the different approaches to construct credibility or confidence intervals for the treatment effect $\mu$ introduced in Section~\ref{sec:meth}, we conducted an extensive simulation study. The simulation scenarios are similar to those considered in \citet{ChungEtAl2013a} and \citet{BrockwellGordon2001}, but extend to smaller numbers of studies included in the meta-analyses and to more pronounced between-trial heterogeneity to reflect the particular circumstances frequently encountered in small populations and rare diseases. We generated data according to the model described in Section~\ref{sec:meth}, where we varied the number of studies involved ($k\in\{3,5,10\}$), the true heterogeneity present ($\tau^2\in\{0,0.01,0.02,0.05,0.1,0.2,0.5,1.0,2.0\}$), and where the individual studies' associated standard errors are generated based on draws from a $\chi^2$-distribution as described in \citet{BrockwellGordon2001}. We then investigated the estimation accuracy for heterogeneity~$\tau$ and effect~$\mu$ by logging the resulting estimates (marginal posterior median in case of the Bayesian methods) and the coverage of true values for confidence and credibility intervals. The simulations were carried out with 10,000 replications per scenario.

Since some readers might be more familiar with the $I^2$ measure to capture the extend of heterogeneity than $\tau^2$, we report here median values of the simulated $I^2$ depending on $\tau^2$ for illustrative purposes. For instance, $\tau^2=0.02, 0.10, 0.5, 2$ resulted in median $I^2$ of 0.11, 0.37, 0.75, and 0.92.

\begin{figure}
\begin{center}
\rotatebox{90}{\hspace{0.9cm} Frequentist}
\includegraphics[width=0.3\linewidth]{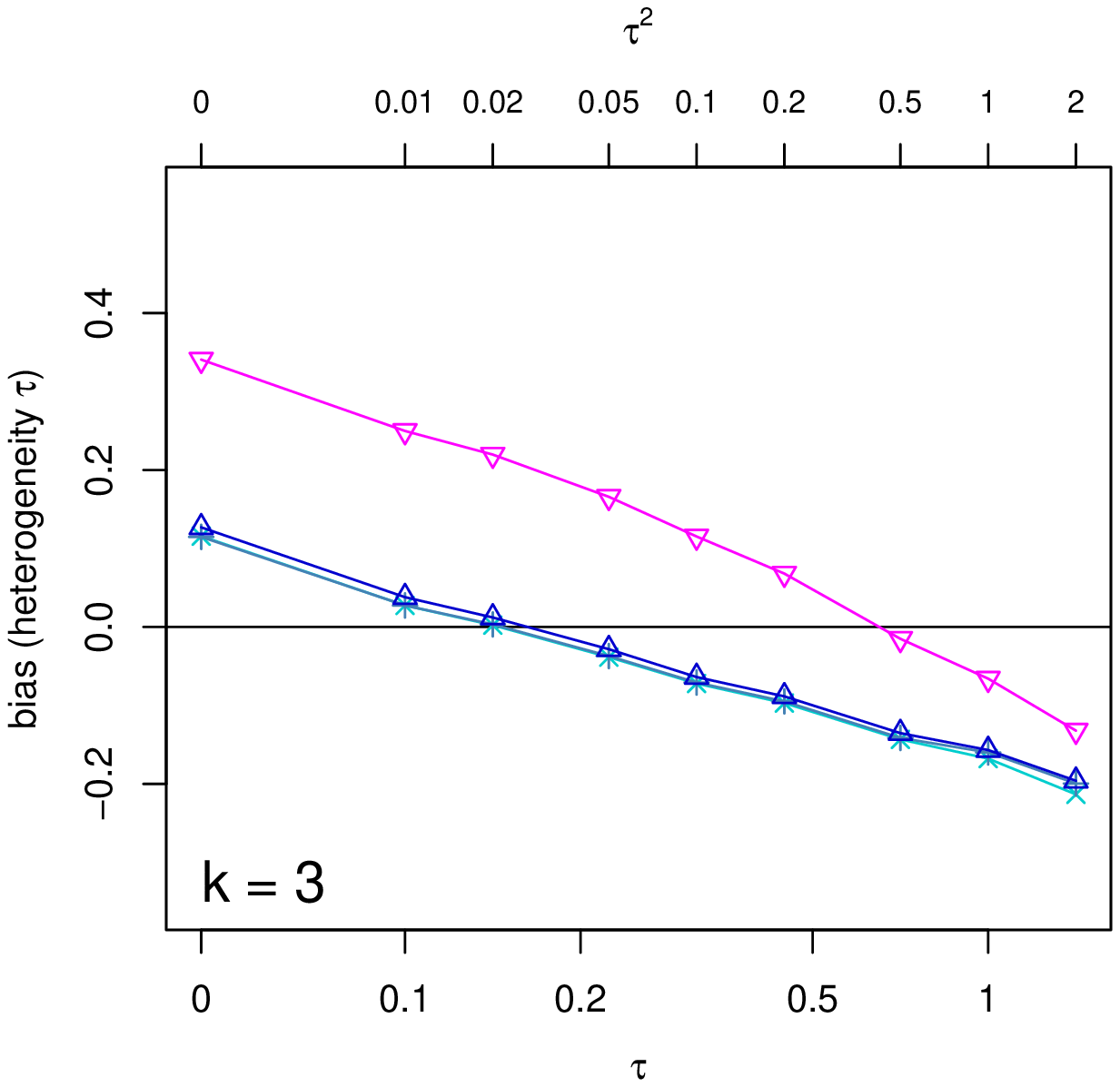}
\includegraphics[width=0.3\linewidth]{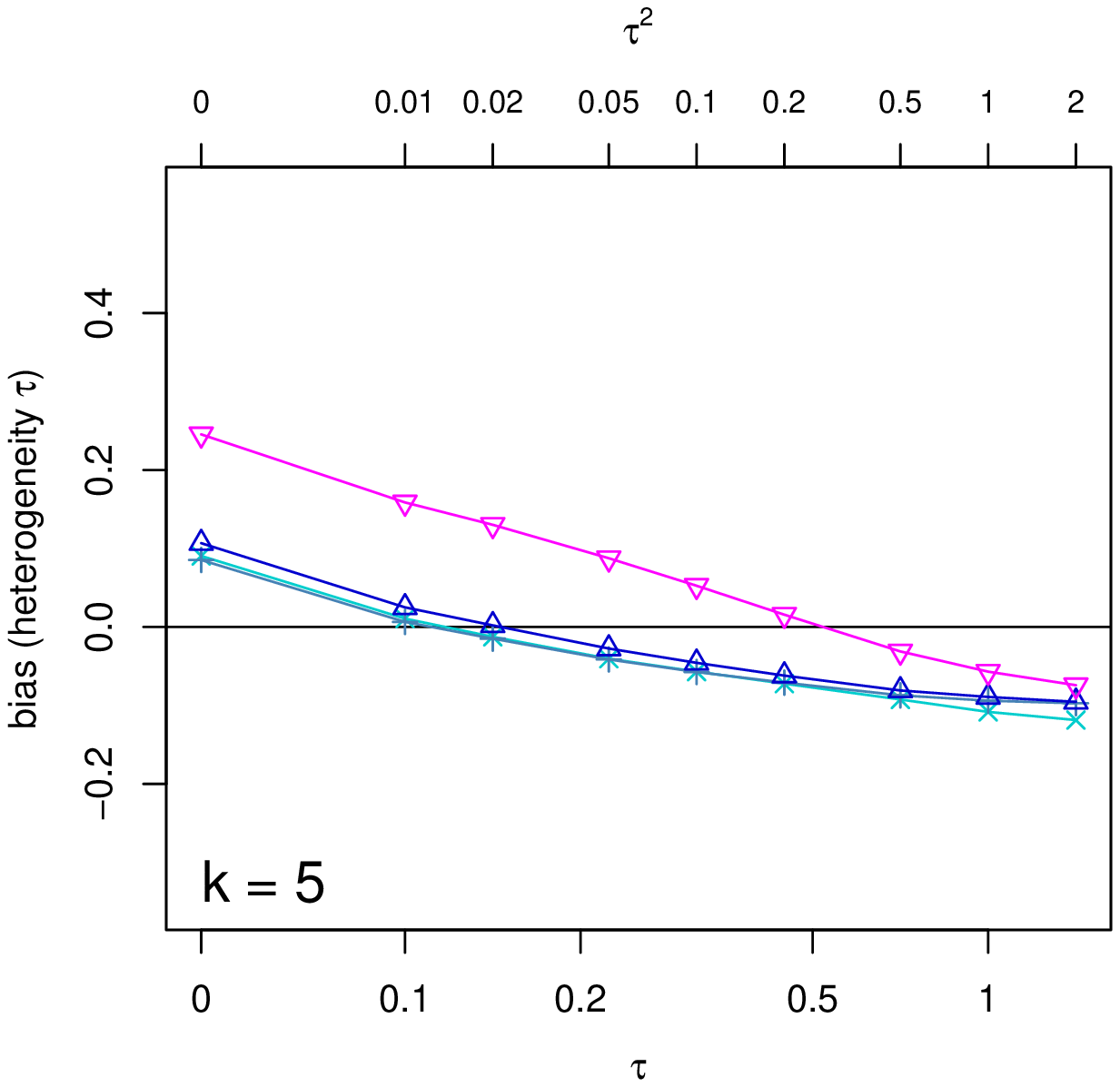} 
\includegraphics[width=0.3\linewidth]{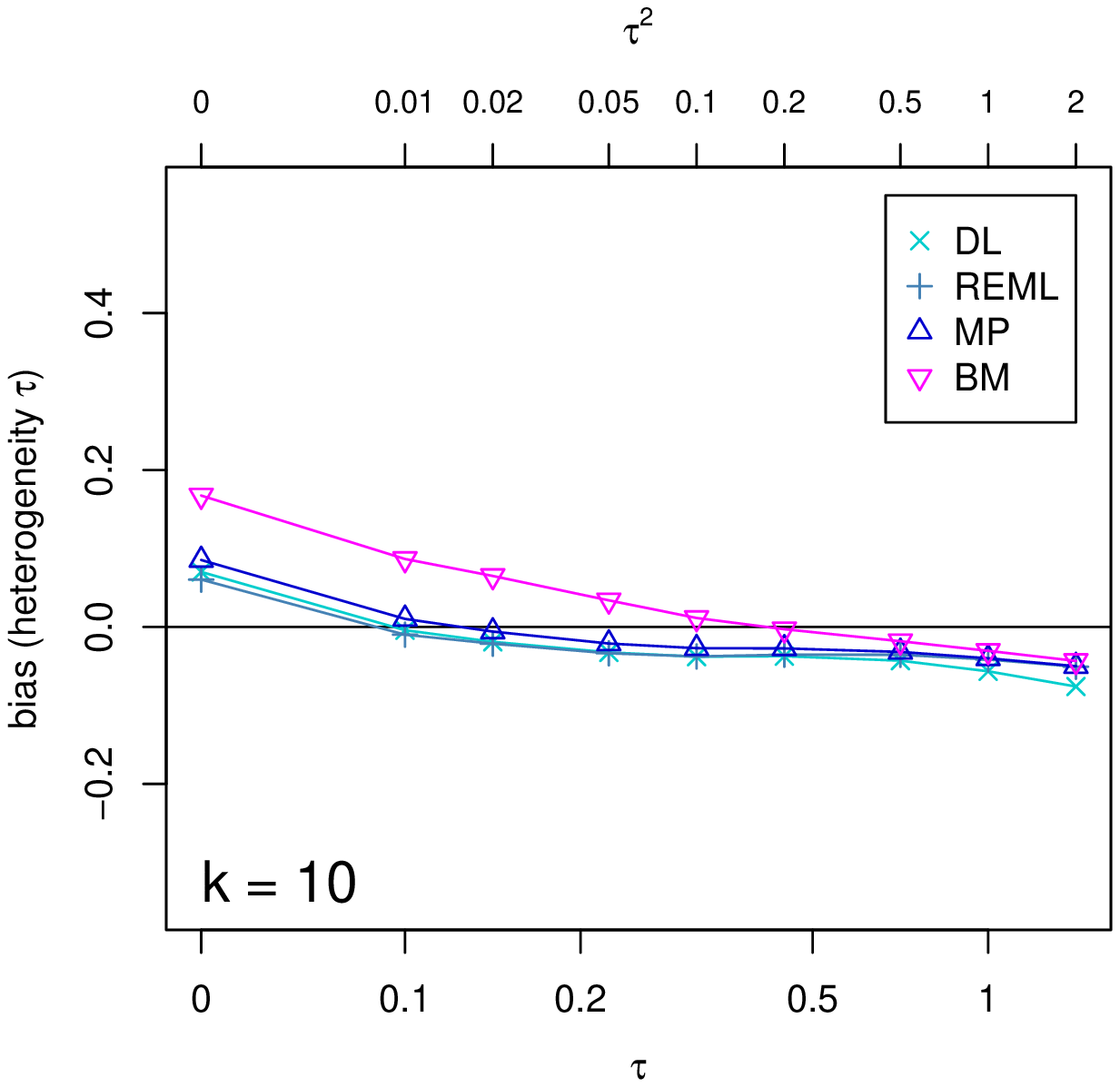} \\
\rotatebox{90}{\hspace{1.3cm} Bayes}
\includegraphics[width=0.3\linewidth]{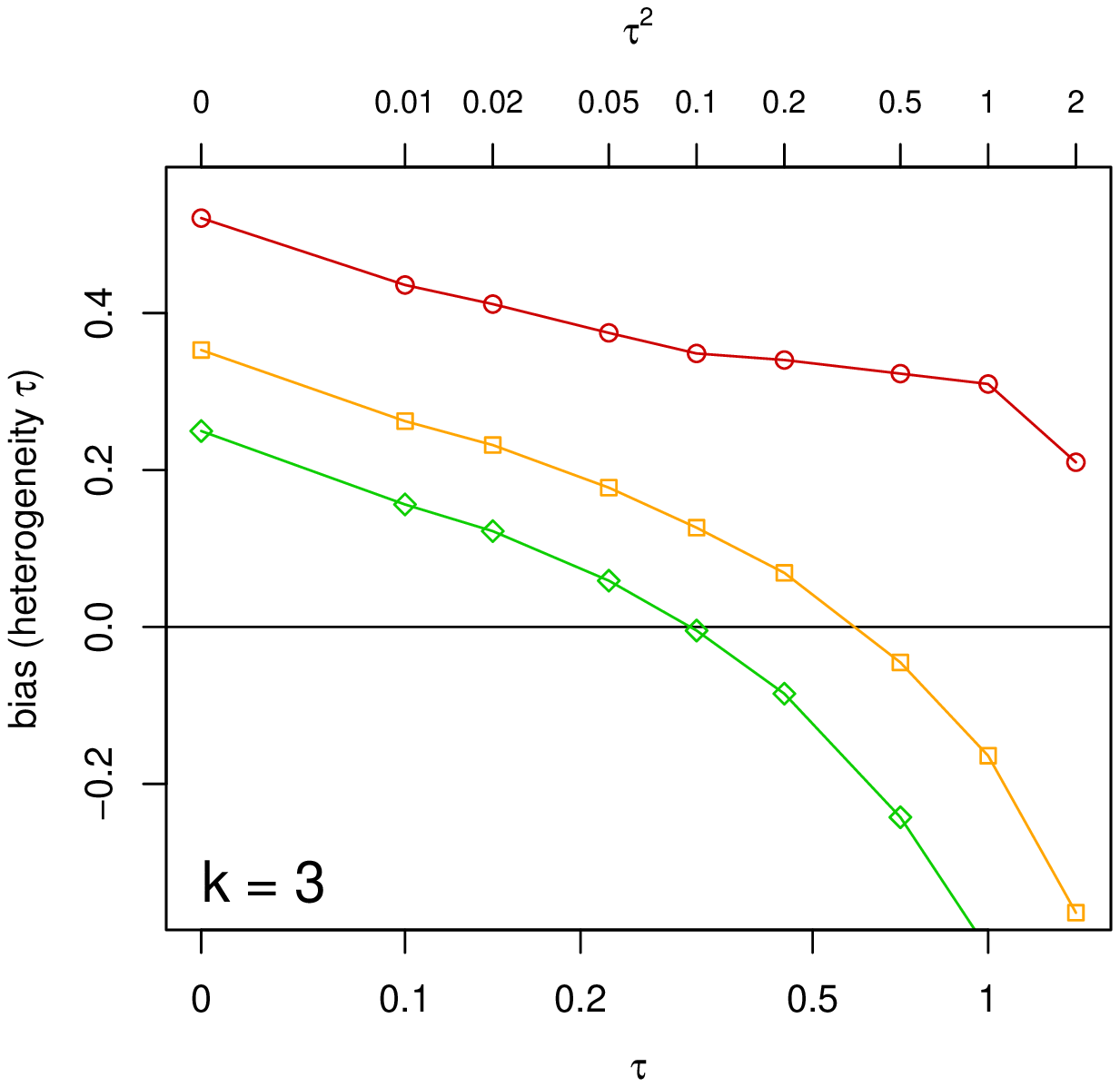}
\includegraphics[width=0.3\linewidth]{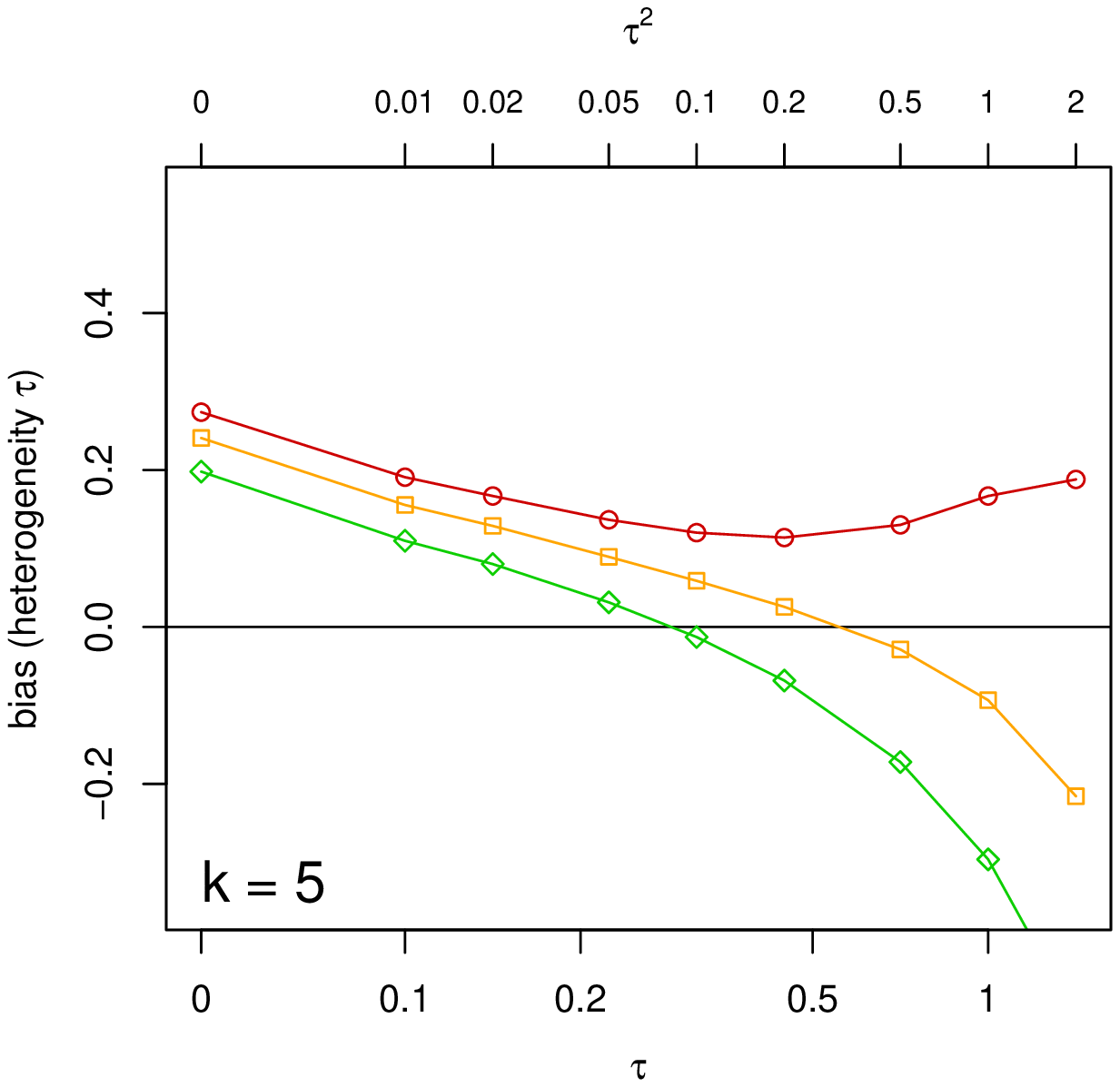} 
\includegraphics[width=0.3\linewidth]{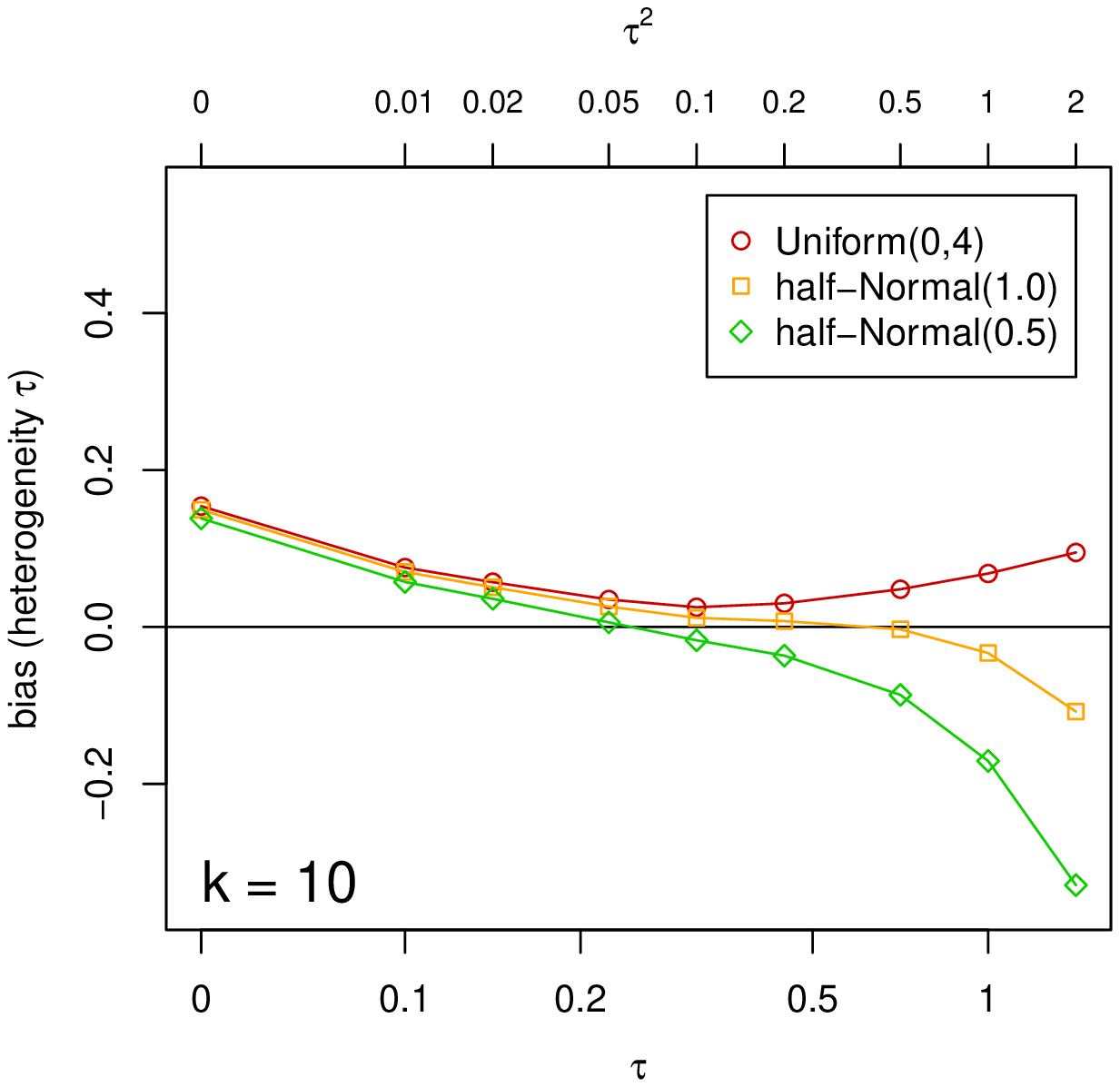} 
\caption{Bias in estimating the between-study heterogeneity $\tau$ for various estimators and for several numbers $k$ of studies included in the meta-analyses.} \label{fig:bias}
\end{center}
\end{figure}

Figure~\ref{fig:bias} shows the bias of several estimators of $\tau$, given various numbers of studies $k$ included in the meta-analyses and a range of true values of $\tau$. The extent of bias strongly depends on the number of studies $k$ for all estimators considered with small bias for large $k$ and substantial bias in situations typically encountered in small populations research. Here the direction and size of bias largely depends on the value of the $\tau$ itself and the estimator. For small $\tau$ the bias can naturally be only positive. For medium to large between-trial heterogeneity the likelihood estimators and the \textsl{DL} estimator tend to underestimate~$\tau$ whereas the Bayesian methods tend to overestimate the between-trial heterogeneity. For very large $\tau$ the picture changes again with some of the Bayesian methods leading to estimates which are in expectation well below the true value depending on the choice of prior. For large and very large $\tau$ the likelihood based estimators and the \textsl{DL} display substantial negative bias. 

\begin{figure}
\begin{center}
\includegraphics[width=0.32\linewidth]{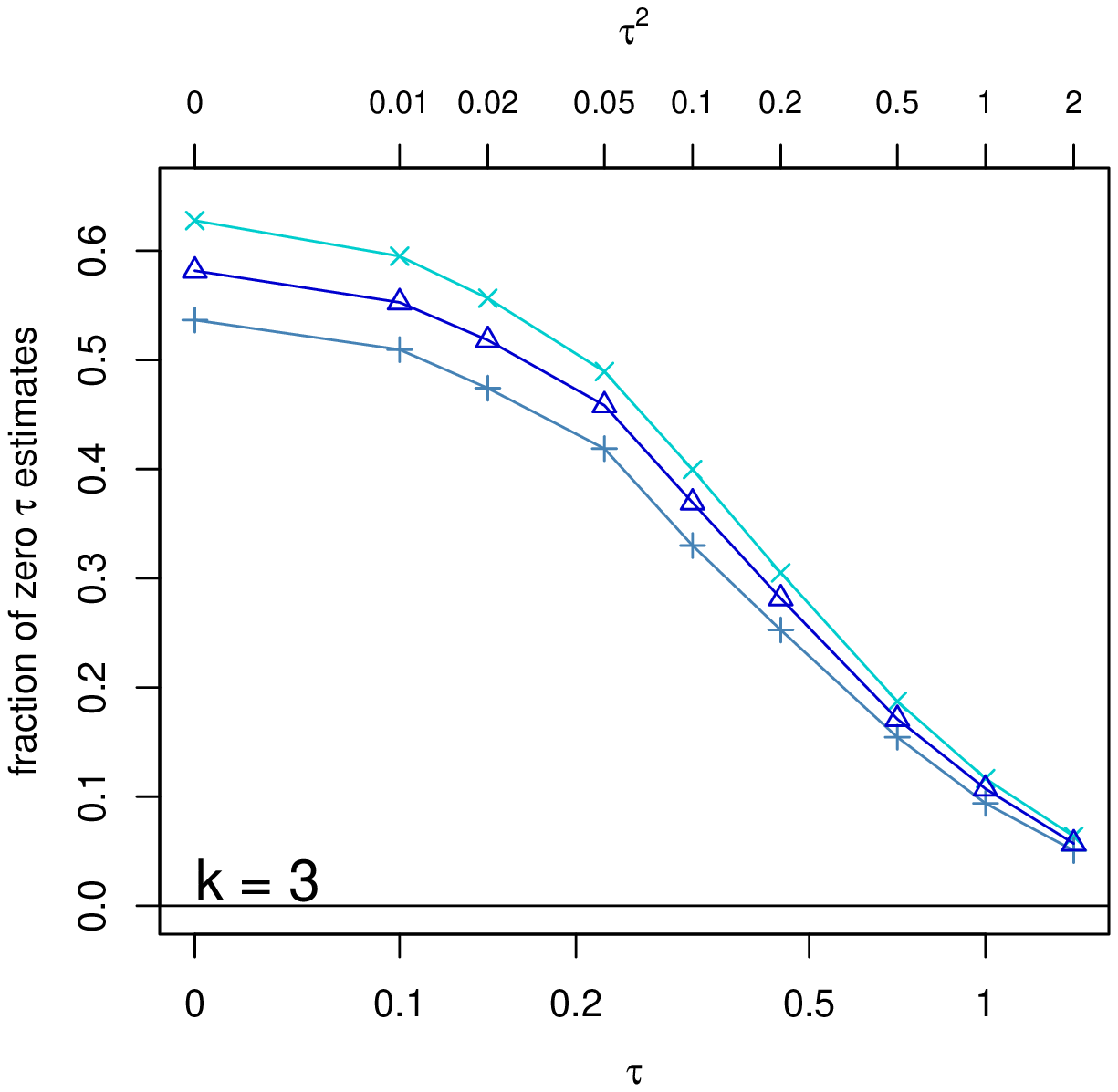}
\includegraphics[width=0.32\linewidth]{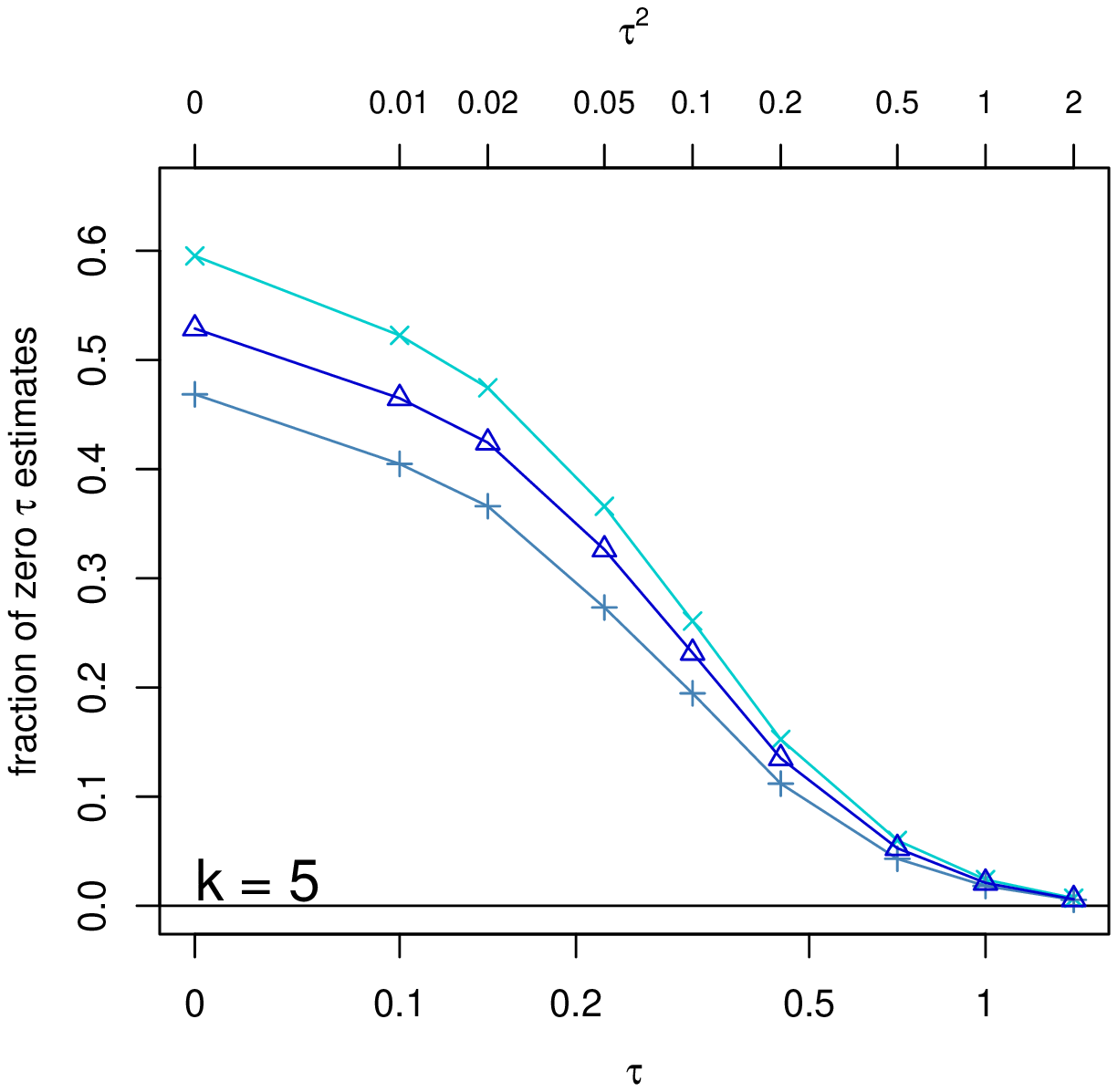}
\includegraphics[width=0.32\linewidth]{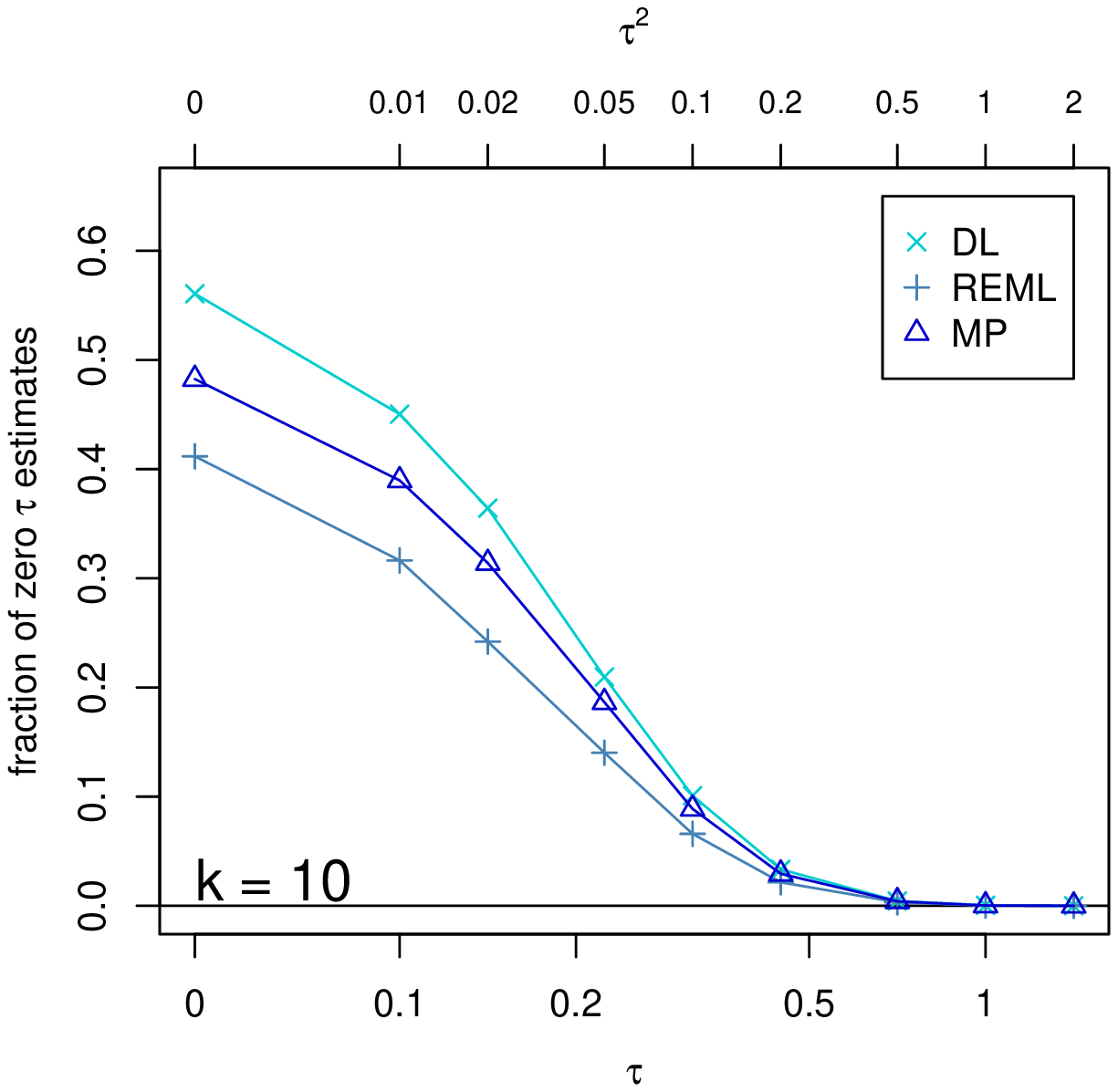}
\caption{Proportion of estimates of the between-study heterogeneity $\tau$ equal to zero for those estimators that are not strictly positive by construction depending on the number $k$ of studies included in the meta-analyses.} \label{fig:tau0}
\end{center}
\end{figure}

For those estimators that are not strictly positive by construction (i.e. \textsl{DL}, \textsl{REML} and \textsl{MP}), Figure~\ref{fig:tau0} shows the proportion of estimates of the between-study heterogeneity $\tau$ equal to zero depending on the number $k$ of studies included in the meta-analyses. In particular for small $k$ the proportion of estimates being equal to 0 is substantial for small to moderate $\tau$. This effect diminishes with increasing $k$. As for the bias, the differences between the three estimators are rather small. 

Given the differences between the seven estimators of $\tau$ one might expect that this would result in differences in estimating $\mu$. Interestingly, we did not observe any marked differences in the root mean squared error for the corresponding estimators of the treatment effect $\mu$ with all differences being below 9\% and discrepancies vanishing with increasing $\tau$ and larger $k$ (data not shown).

\begin{figure}
\begin{center}
\rotatebox{90}{\hspace{1.1cm} Normal}
\includegraphics[width=0.3\linewidth]{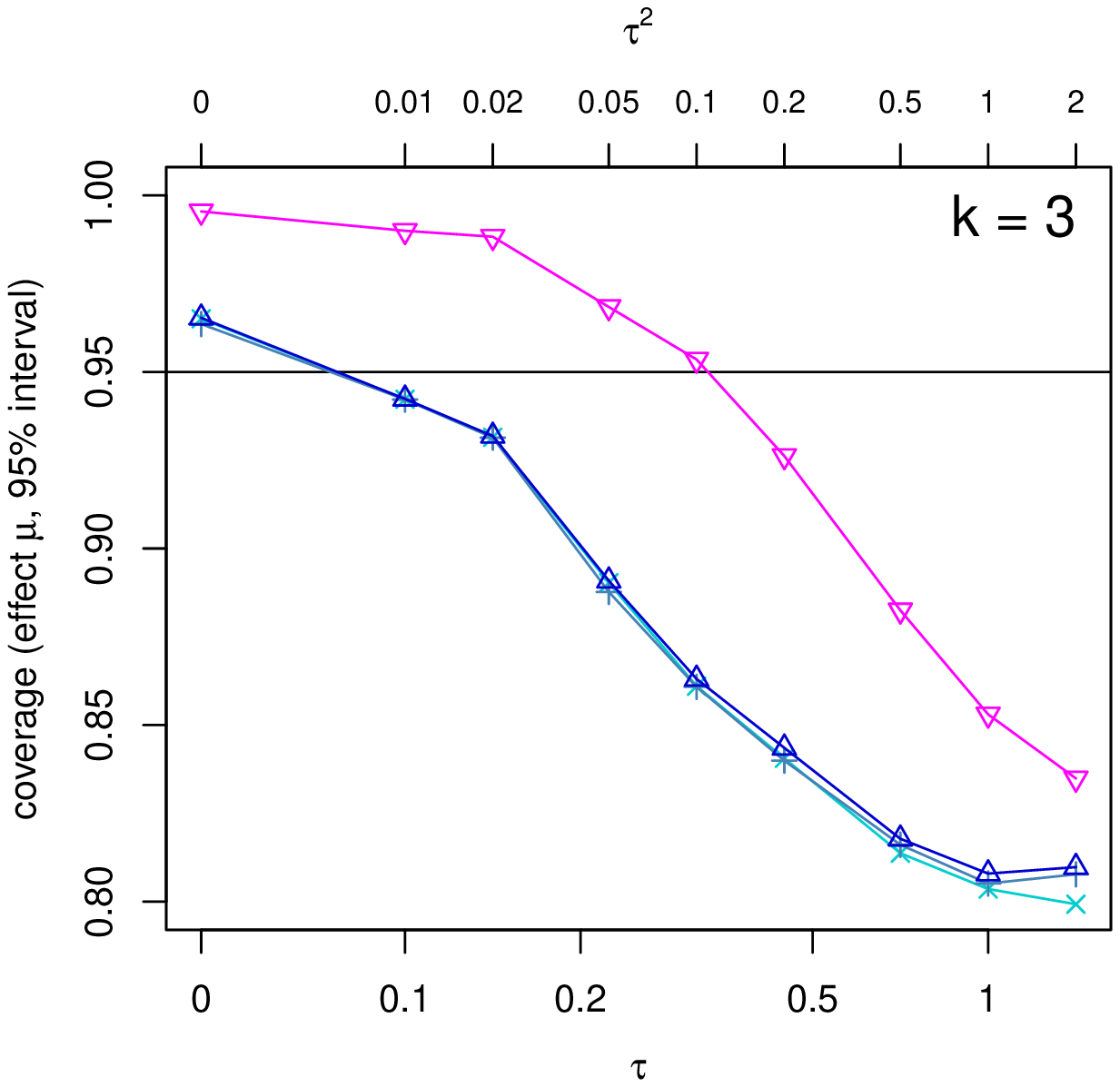} 
\includegraphics[width=0.3\linewidth]{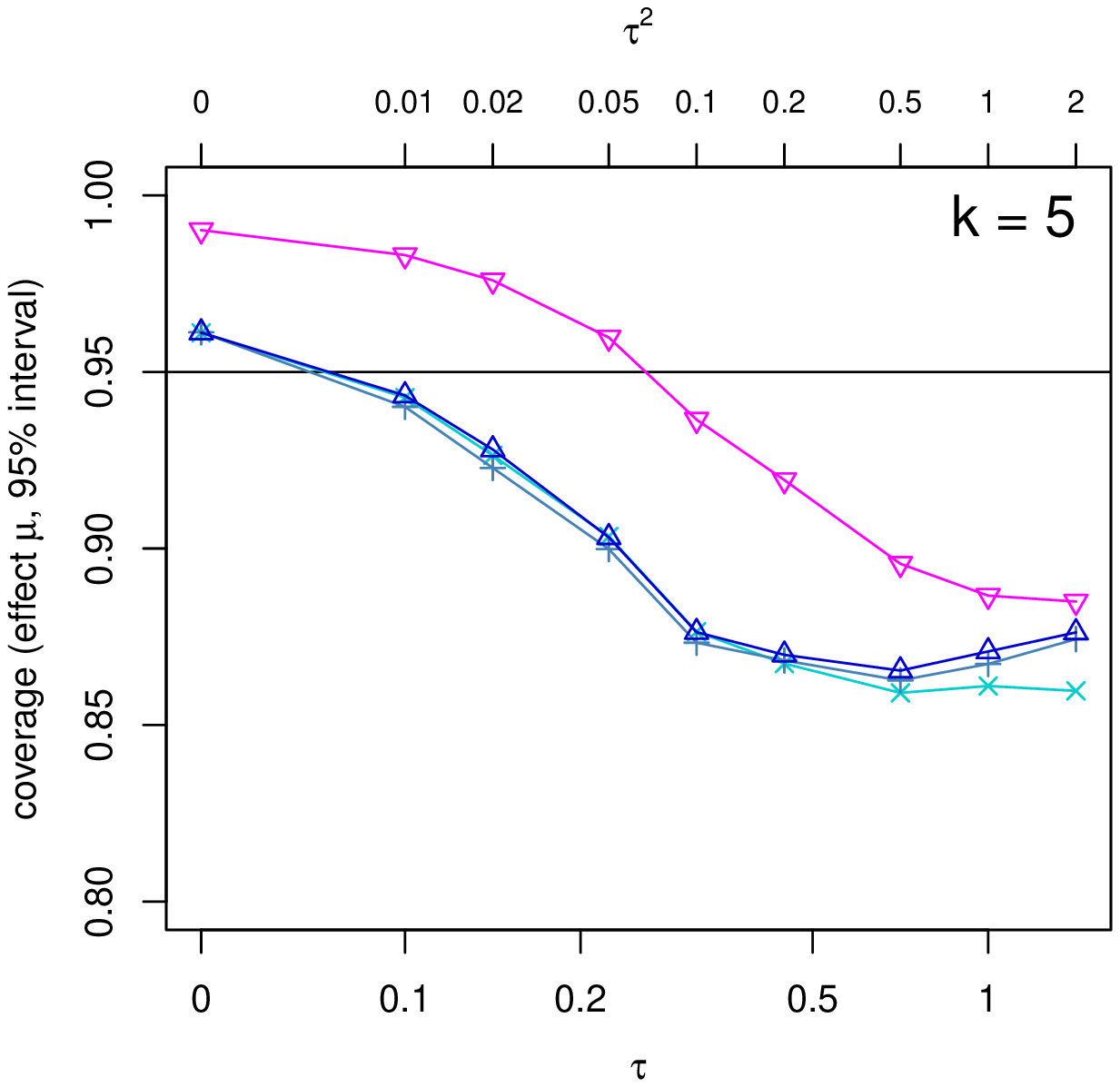}
\includegraphics[width=0.3\linewidth]{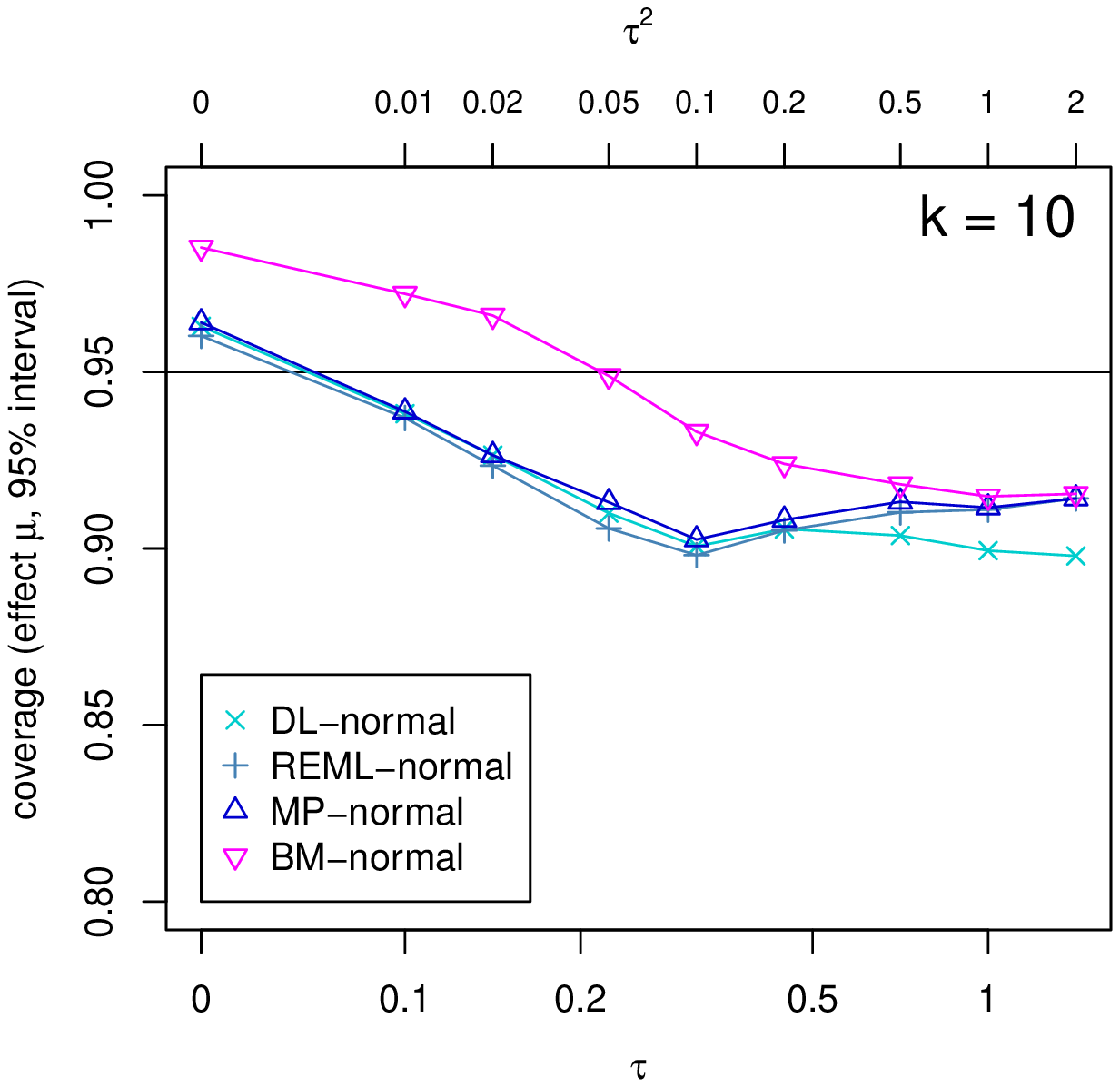} \\
\rotatebox{90}{\hspace{0.5cm} Knapp-Hartung}
\includegraphics[width=0.3\linewidth]{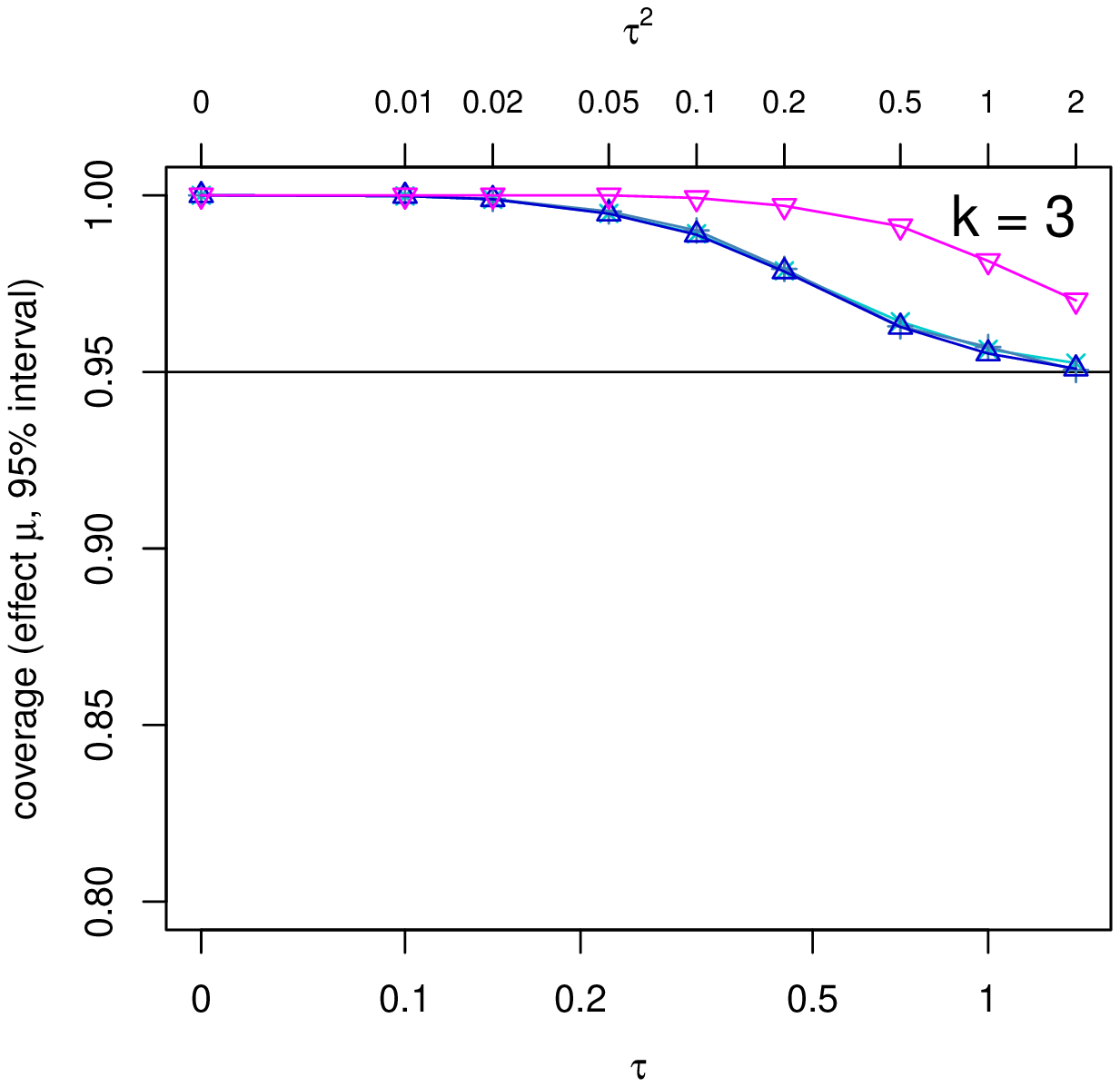} 
\includegraphics[width=0.3\linewidth]{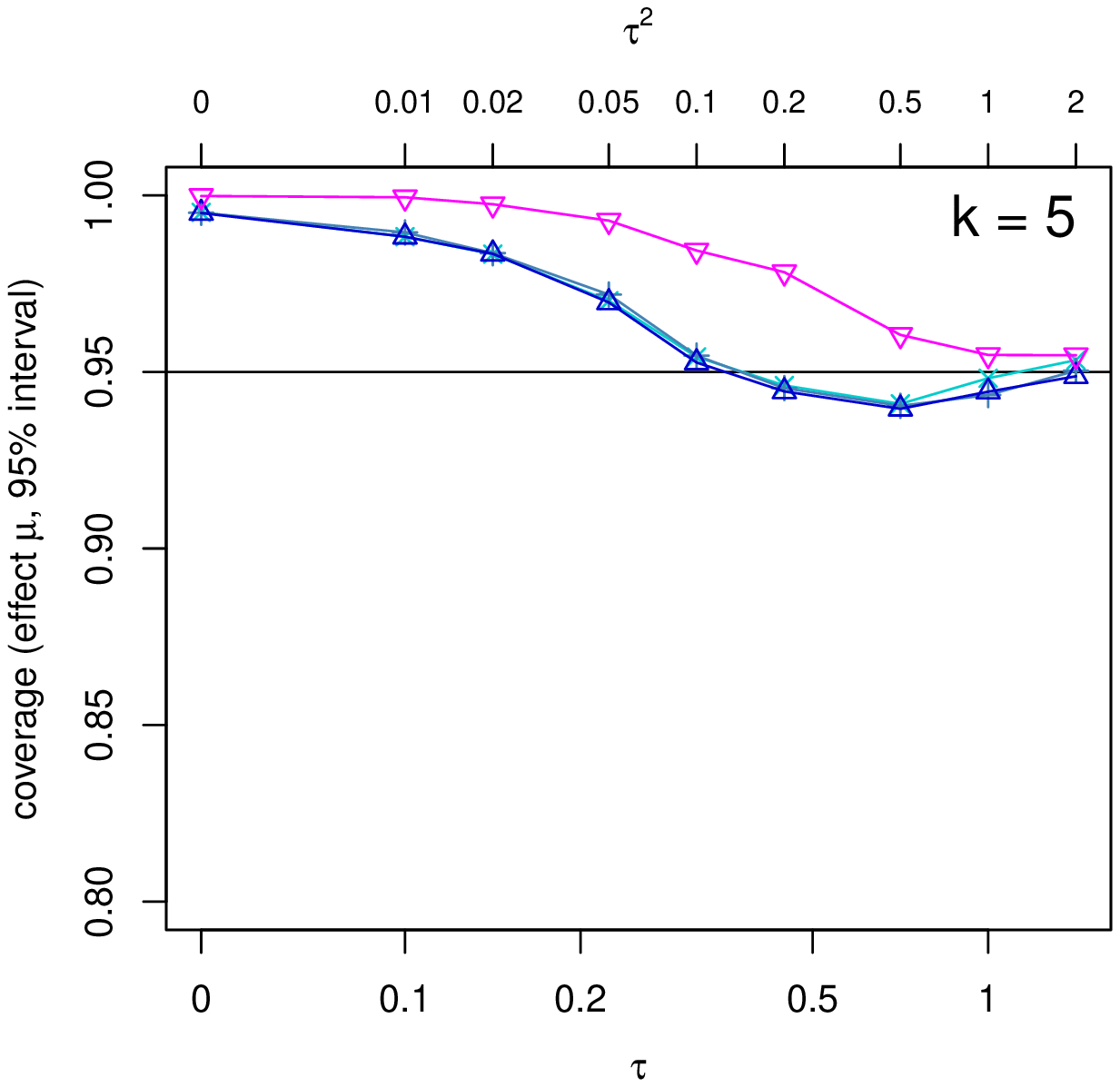}
\includegraphics[width=0.3\linewidth]{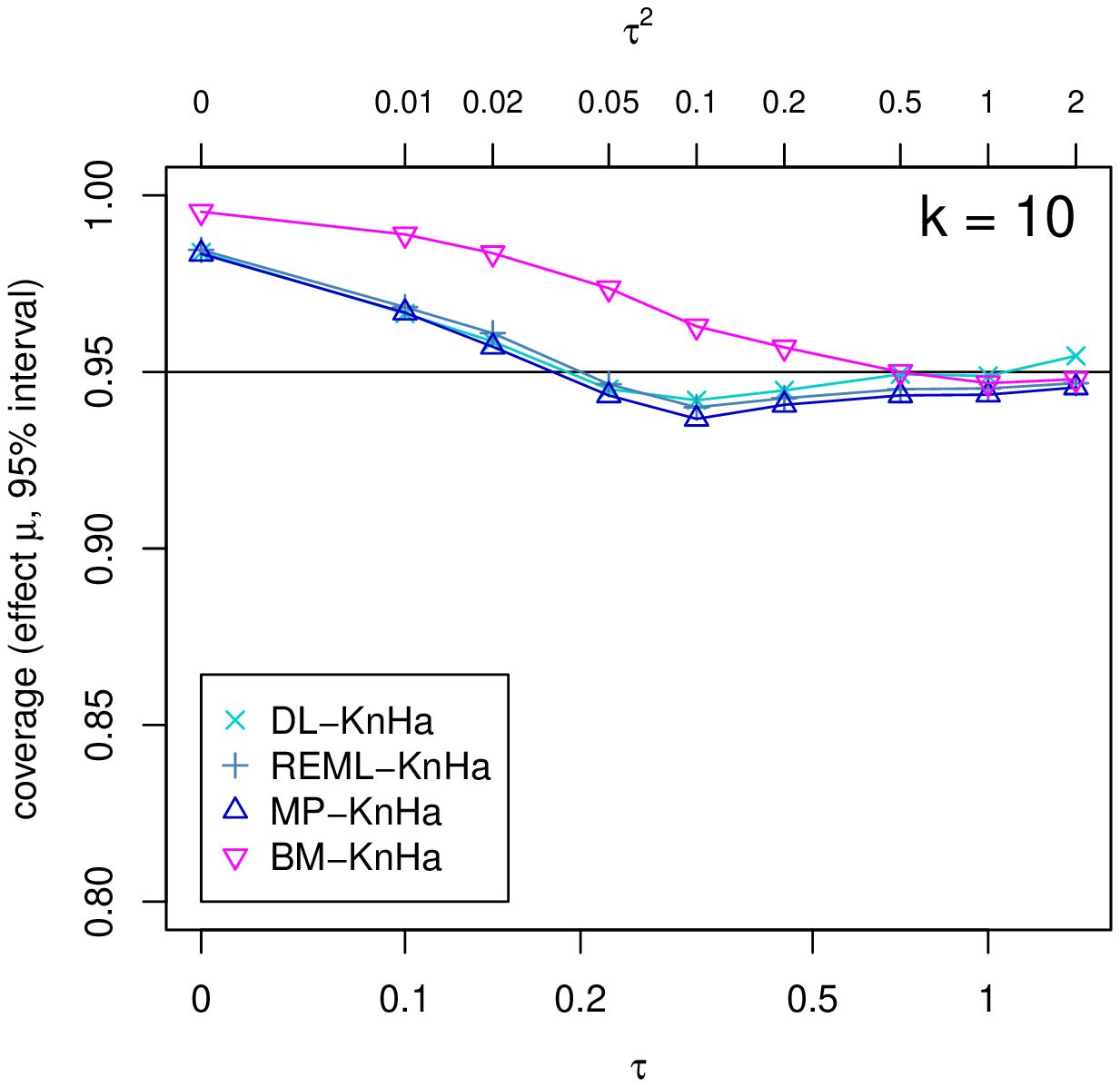} \\
\rotatebox{90}{\hspace{1.3cm} Bayes}
\includegraphics[width=0.3\linewidth]{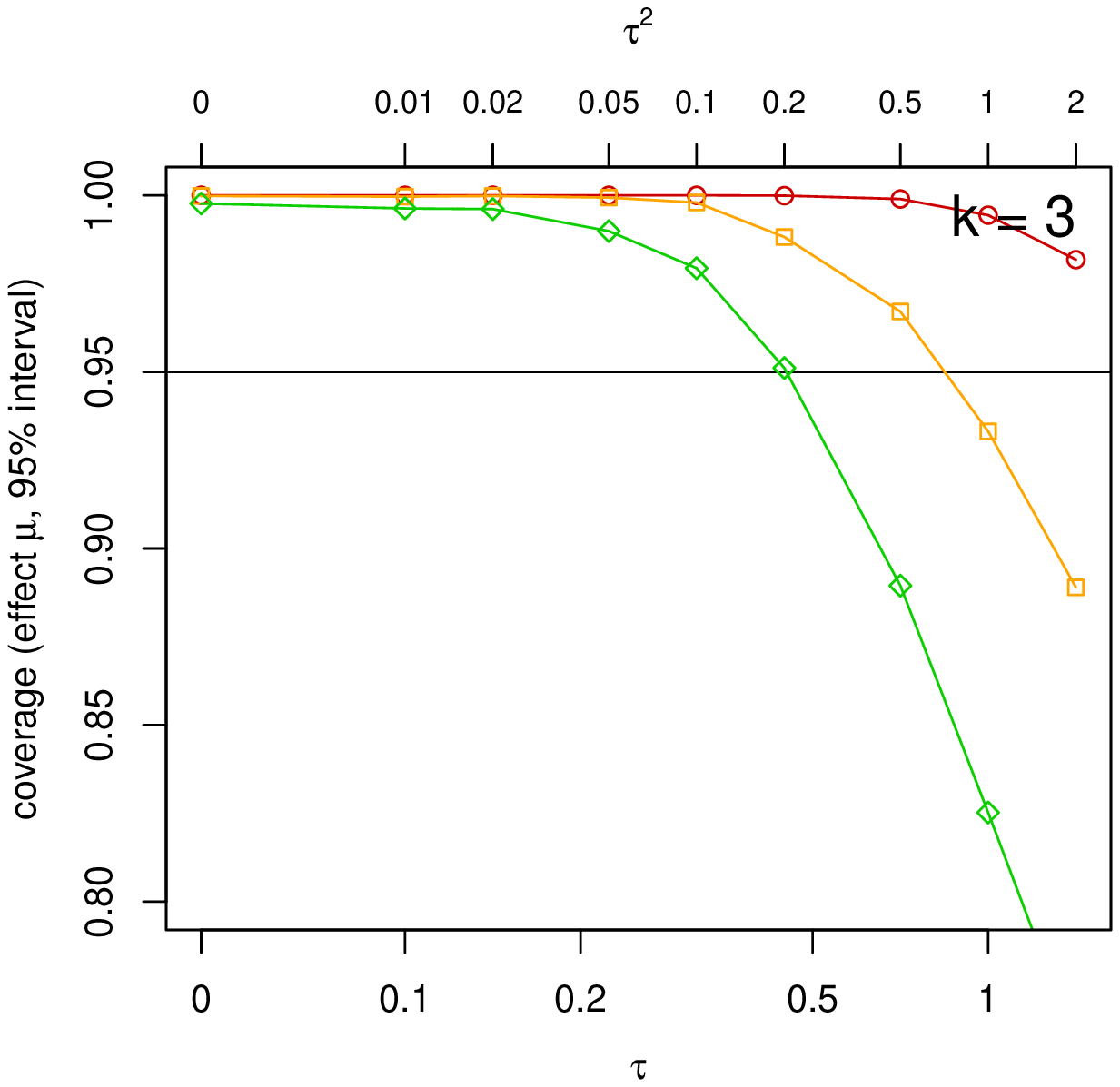} 
\includegraphics[width=0.3\linewidth]{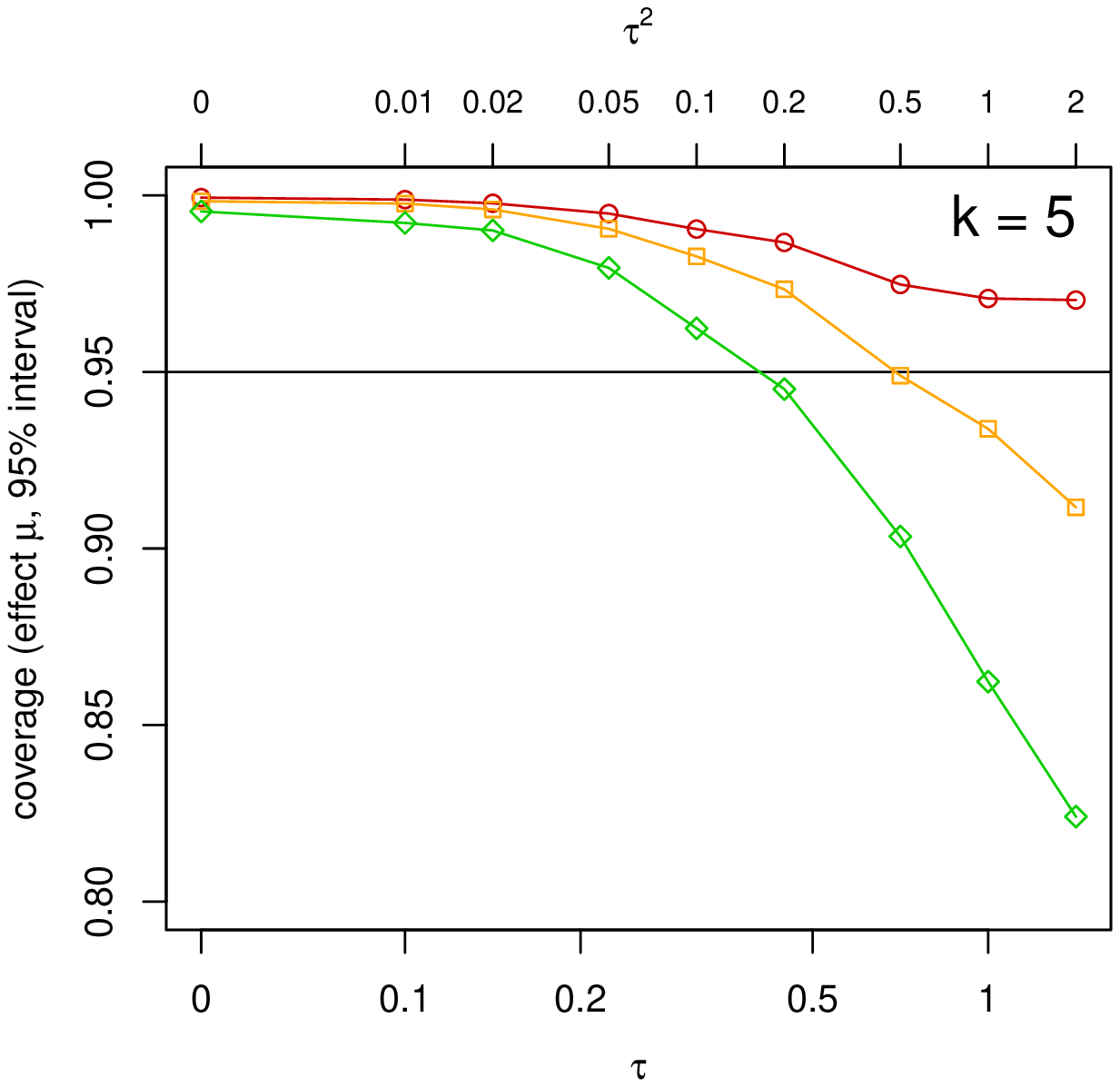}
\includegraphics[width=0.3\linewidth]{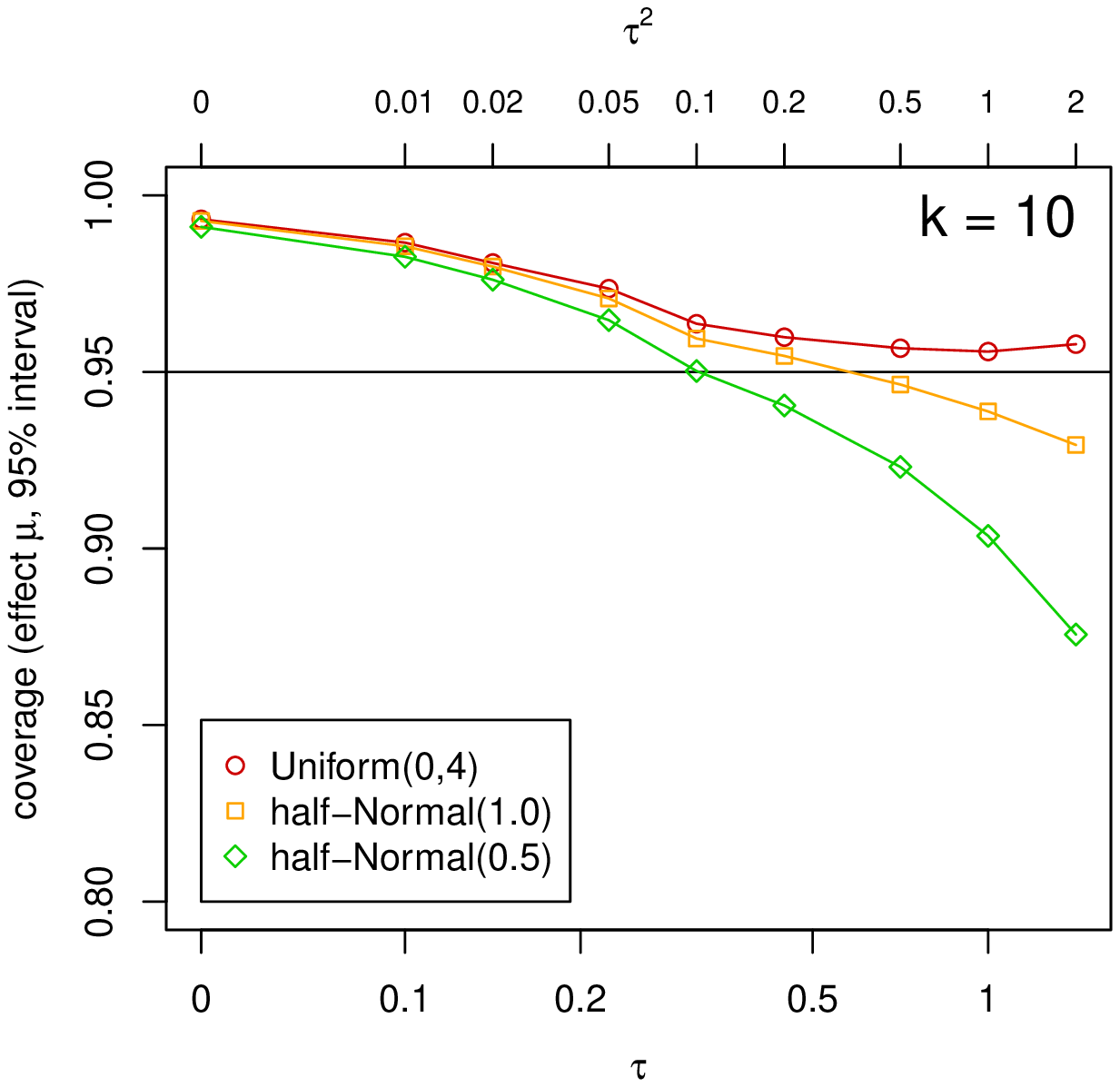}
\caption{Coverage probability for credibility / confidence intervals for the overall treatment effect $\mu$ depending on the between-study heterogeneity $\tau$ and several numbers $k$ of studies included in the meta-analyses using various estimators.} \label{fig:coverage}
\end{center}
\end{figure}

Figure~\ref{fig:coverage} shows the coverage of several credibility and confidence intervals for the treatment effect $\mu$ depending on the between-trial heterogeneity $\tau^2$ and the number of included studies $k$. When considering the bias in estimating $\tau$ above, no estimator outperformed the others over the range of the considered scenarios. Comparing the various approaches with regard to the coverage probabilities of the credibility and confidence intervals for $\mu$, however, the picture becomes much clearer. The frequentist methods based on normal approximation achieve coverage probabilites at or above the nominal level only for relatively small $\tau$ (i.e. $\tau \le 0.1$), for which the Bayesian methods as well as the Knapp-Hartung method are conservative. For most other values of $\tau$ the Bayes and Knapp-Hartung methods outperform the normal approximation approaches in that they provide higher coverage probabilities. However, also for Bayesian and Knapp-Hartung methods the actual coverage can deviate from the nominal confidence level substantially. With appropriate choice of the prior distribution the coverage probabilities of the Bayesian credibility intervals are often closer the the nominal level than the Knapp-Hartung method. The differences between the various methods are not only due to differences in estimating $\tau$, but also reflect differences in the construction of the intervals. By construction the Bayesian credibility intervals are guaranteed to produce coverage probabilities of the nominal level when averaged over the prior distribution. Here we consider coverage probabilities at fixed $\tau$ and therefore expect the coverage probabilities to vary with $\tau$. The frequentist methods considered achieve the coverage probability asymptotically, but their characteristics for finite samples are not obvious. This can be seen in Figure~\ref{fig:coverage} where the coverage probabilities are closer to the nominal level for larger $k$. 

\begin{figure}
\begin{center}
\rotatebox{90}{\hspace{1.1cm} Normal}
\includegraphics[width=0.3\linewidth]{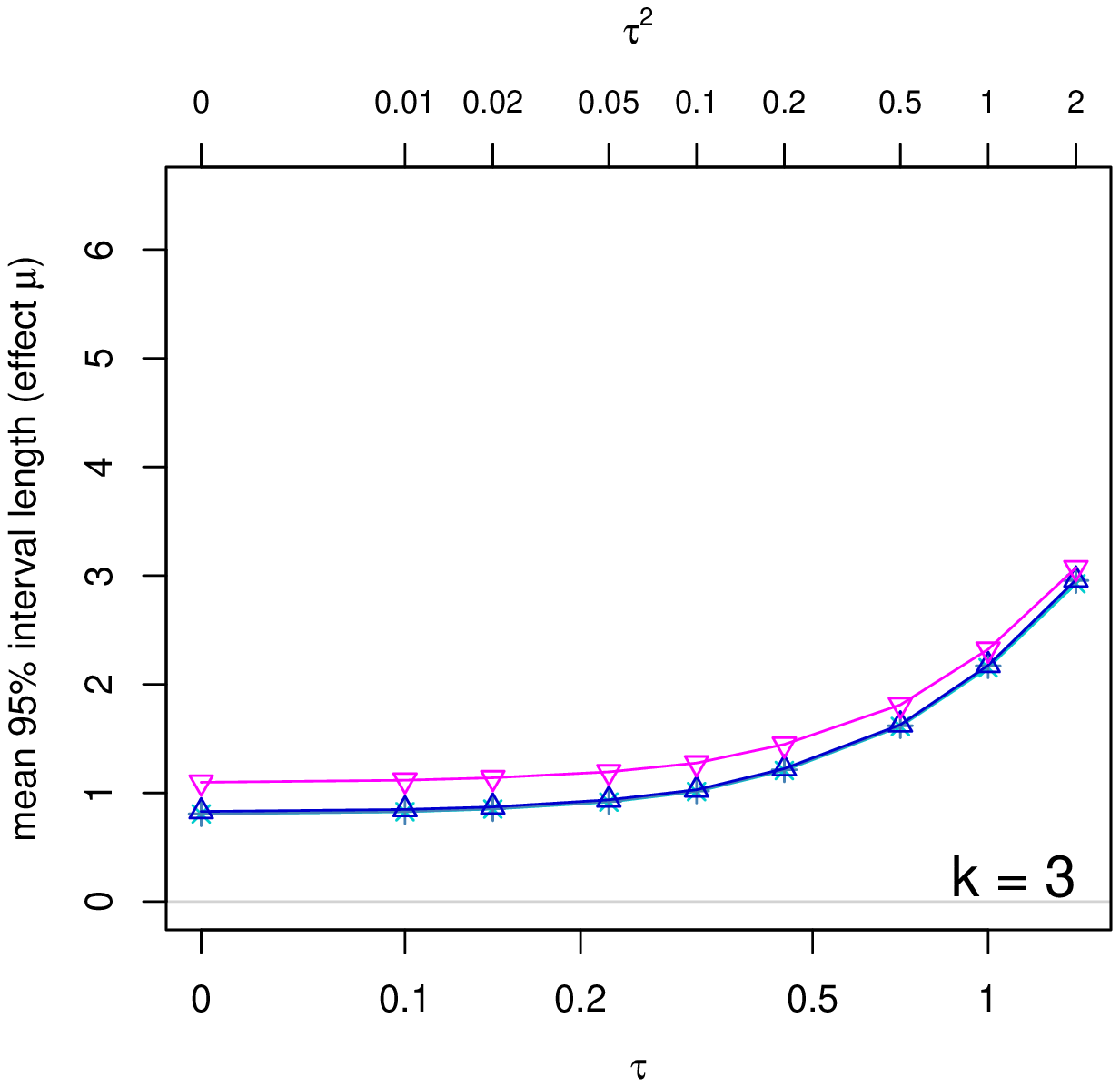} 
\includegraphics[width=0.3\linewidth]{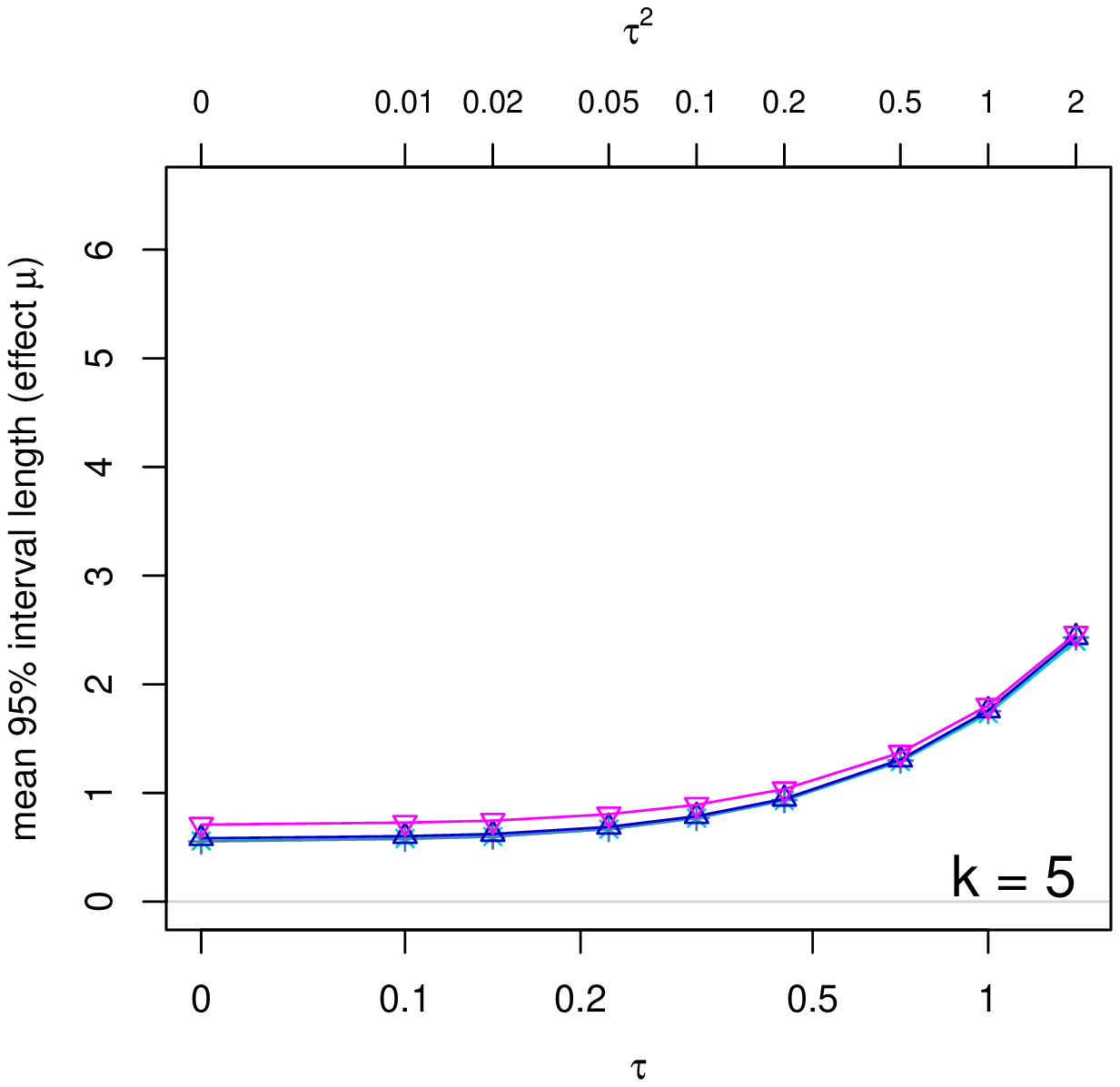} 
\includegraphics[width=0.3\linewidth]{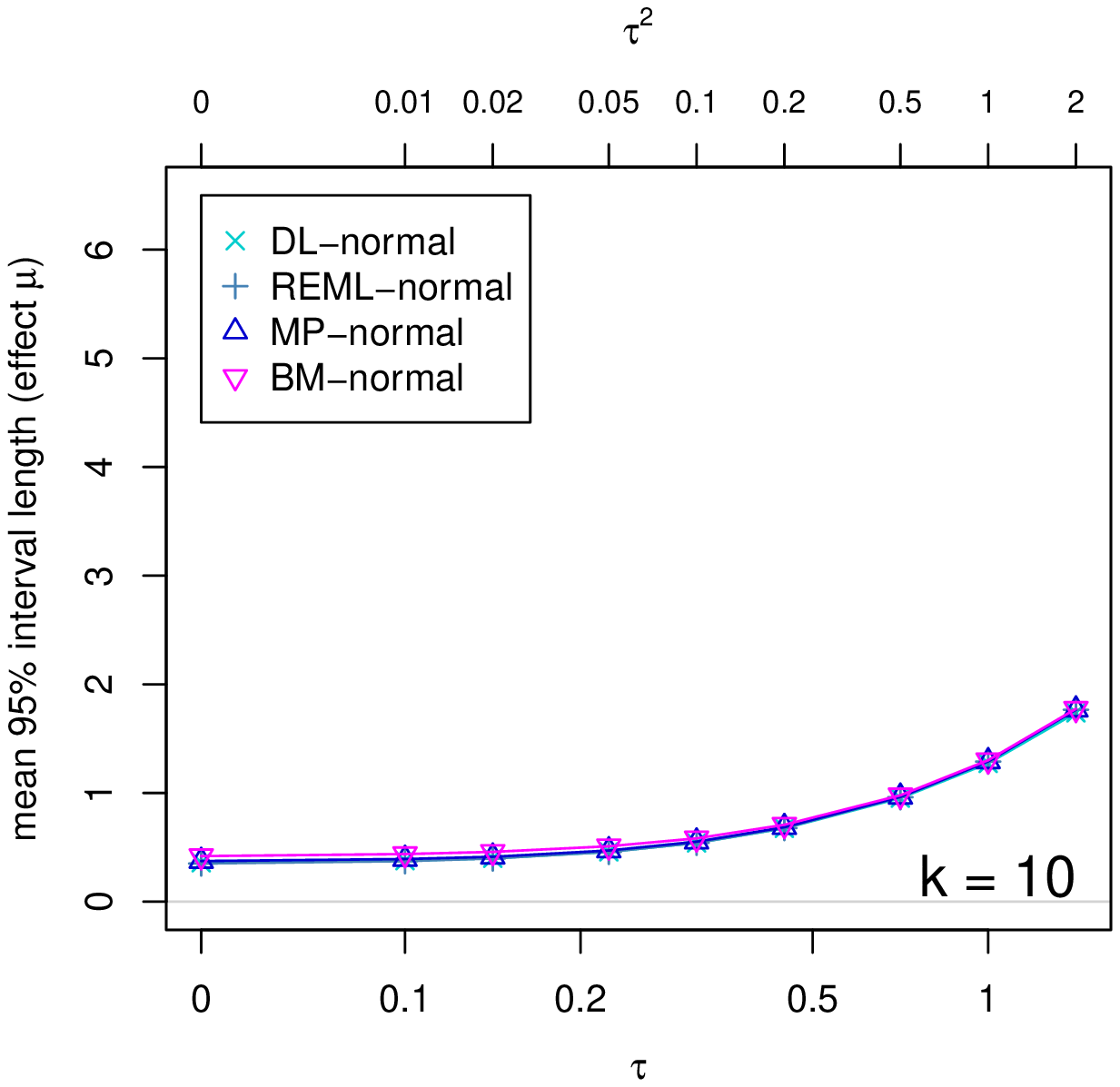} \\
\rotatebox{90}{\hspace{0.5cm} Knapp-Hartung}
\includegraphics[width=0.3\linewidth]{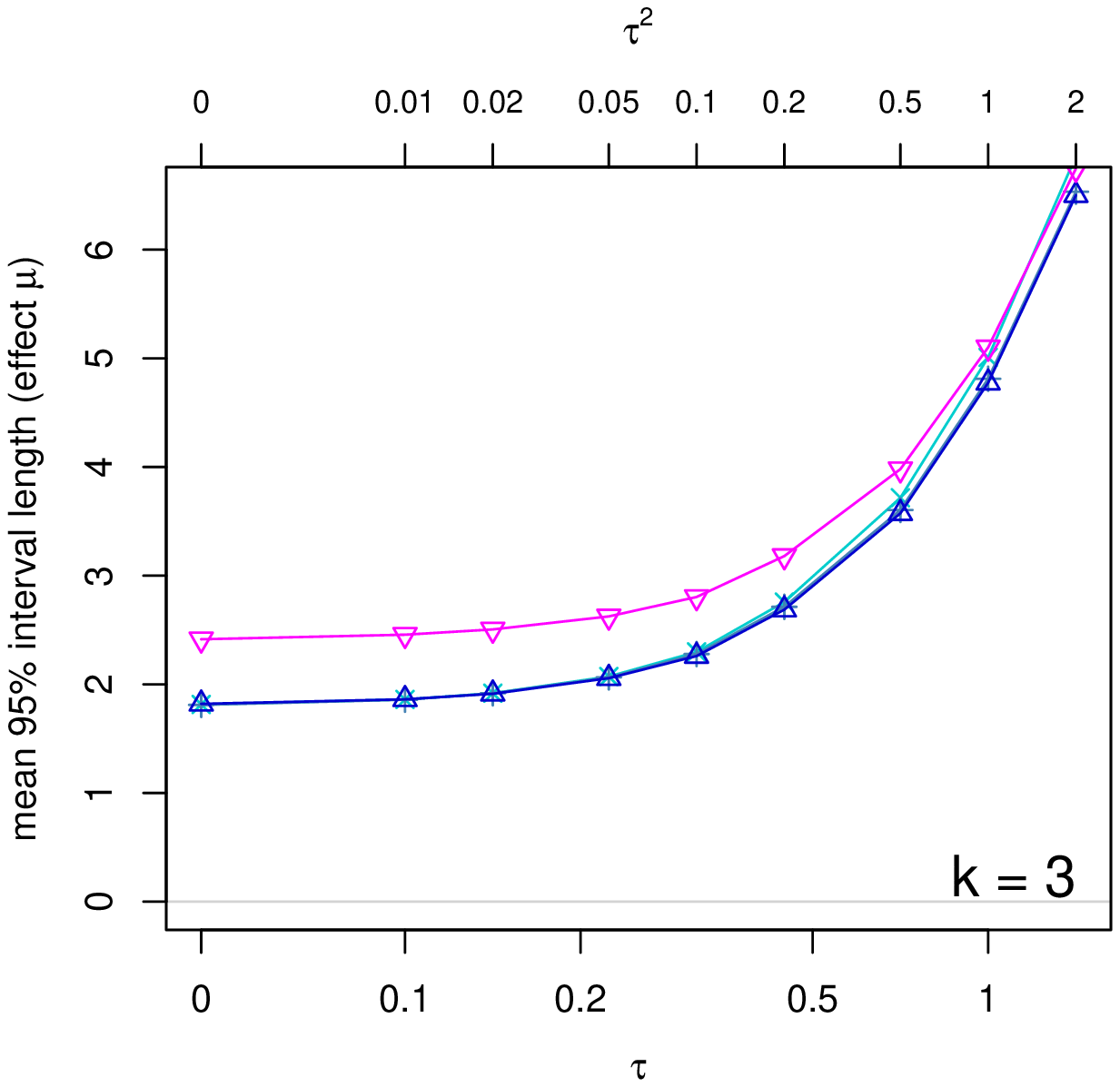} 
\includegraphics[width=0.3\linewidth]{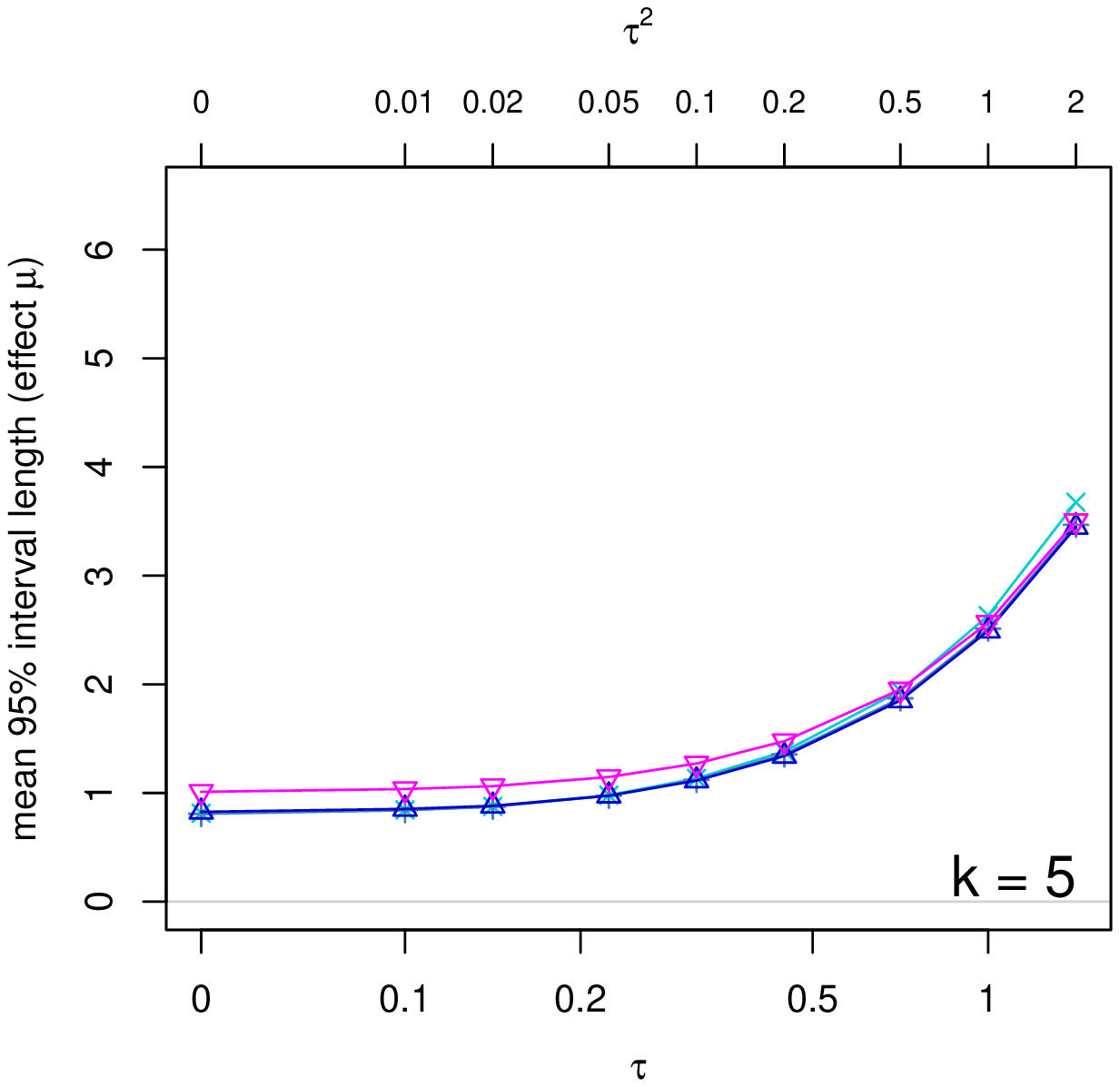} 
\includegraphics[width=0.3\linewidth]{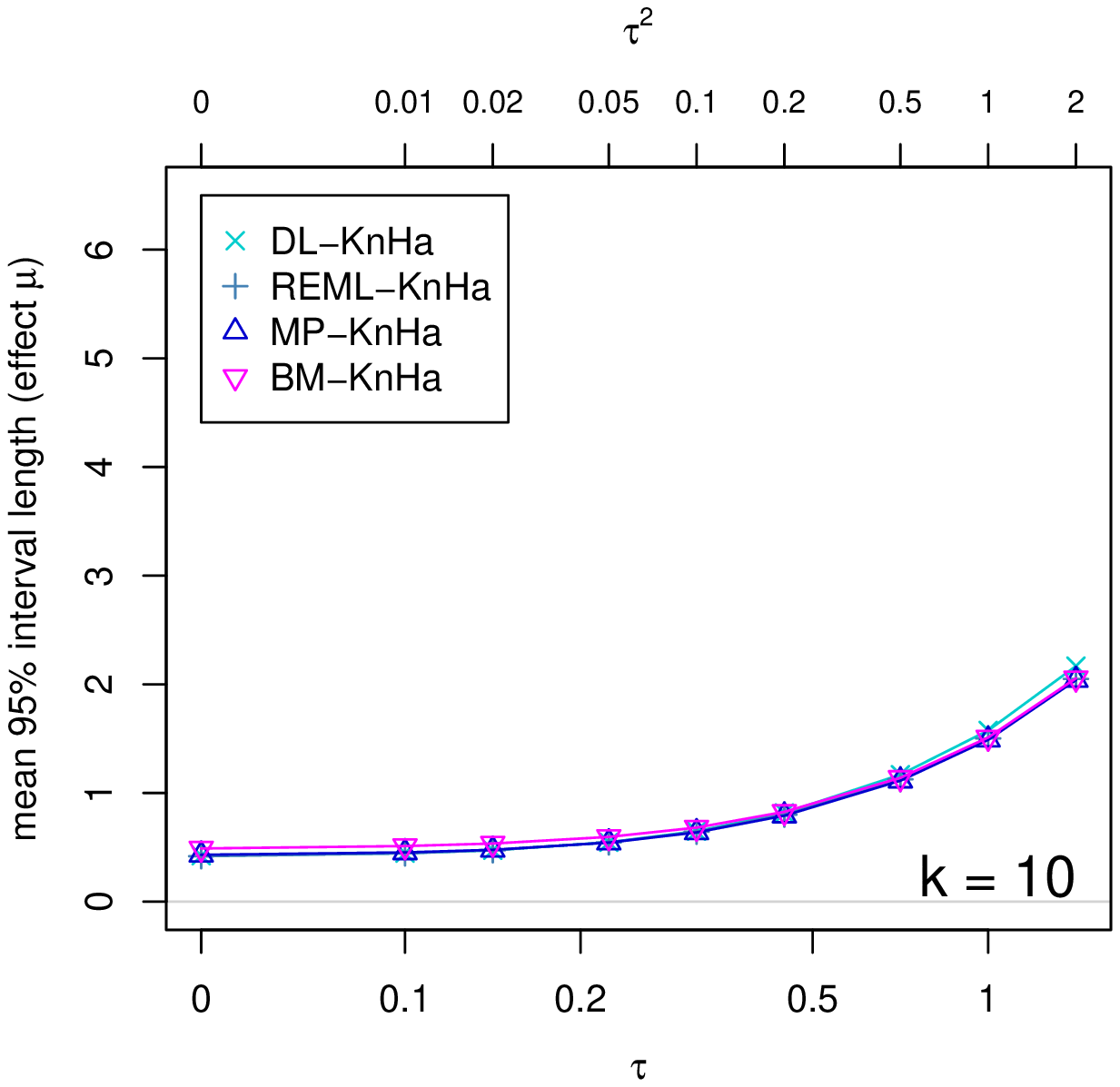} \\
\rotatebox{90}{\hspace{1.3cm} Bayes}
\includegraphics[width=0.3\linewidth]{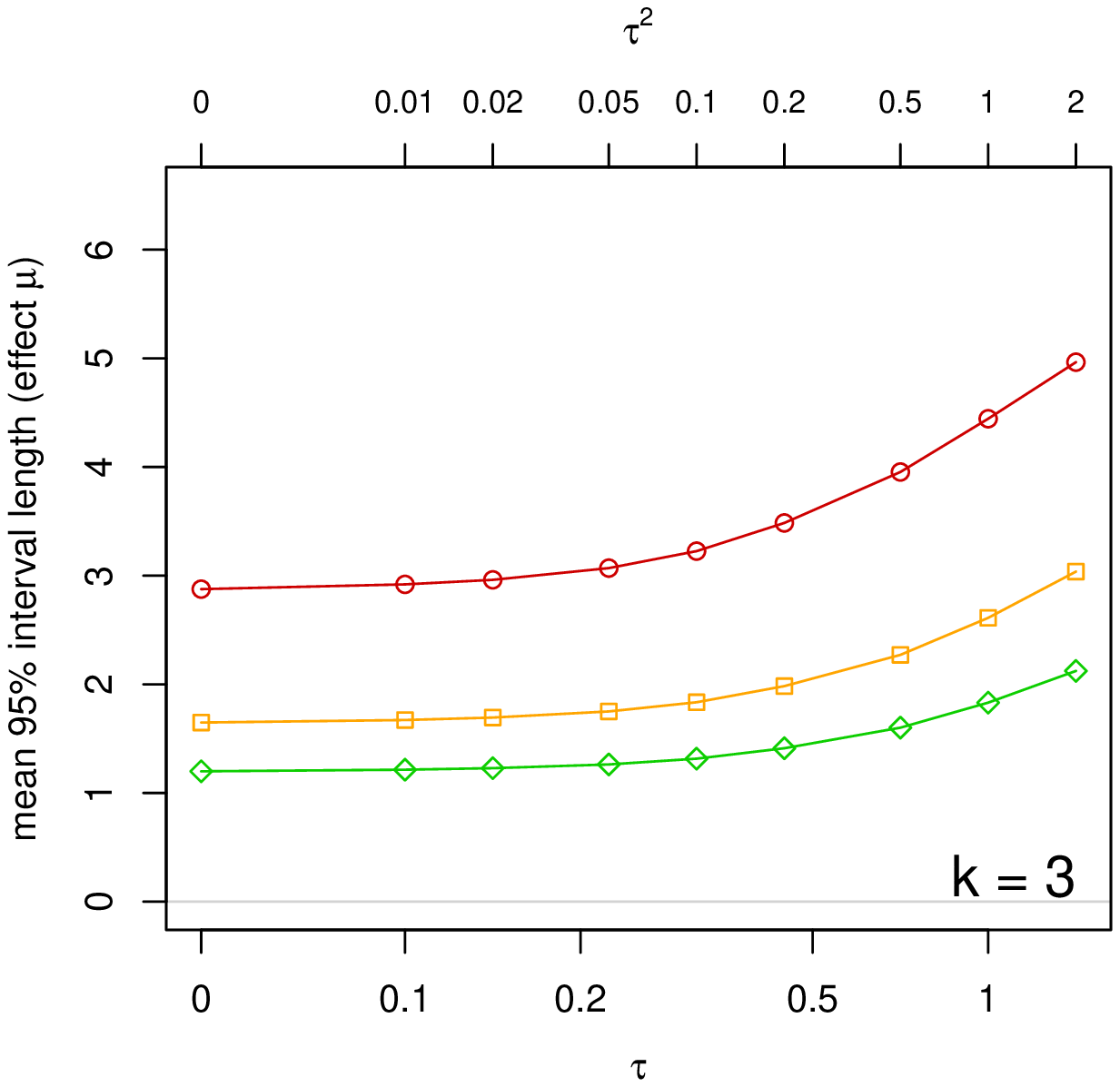} 
\includegraphics[width=0.3\linewidth]{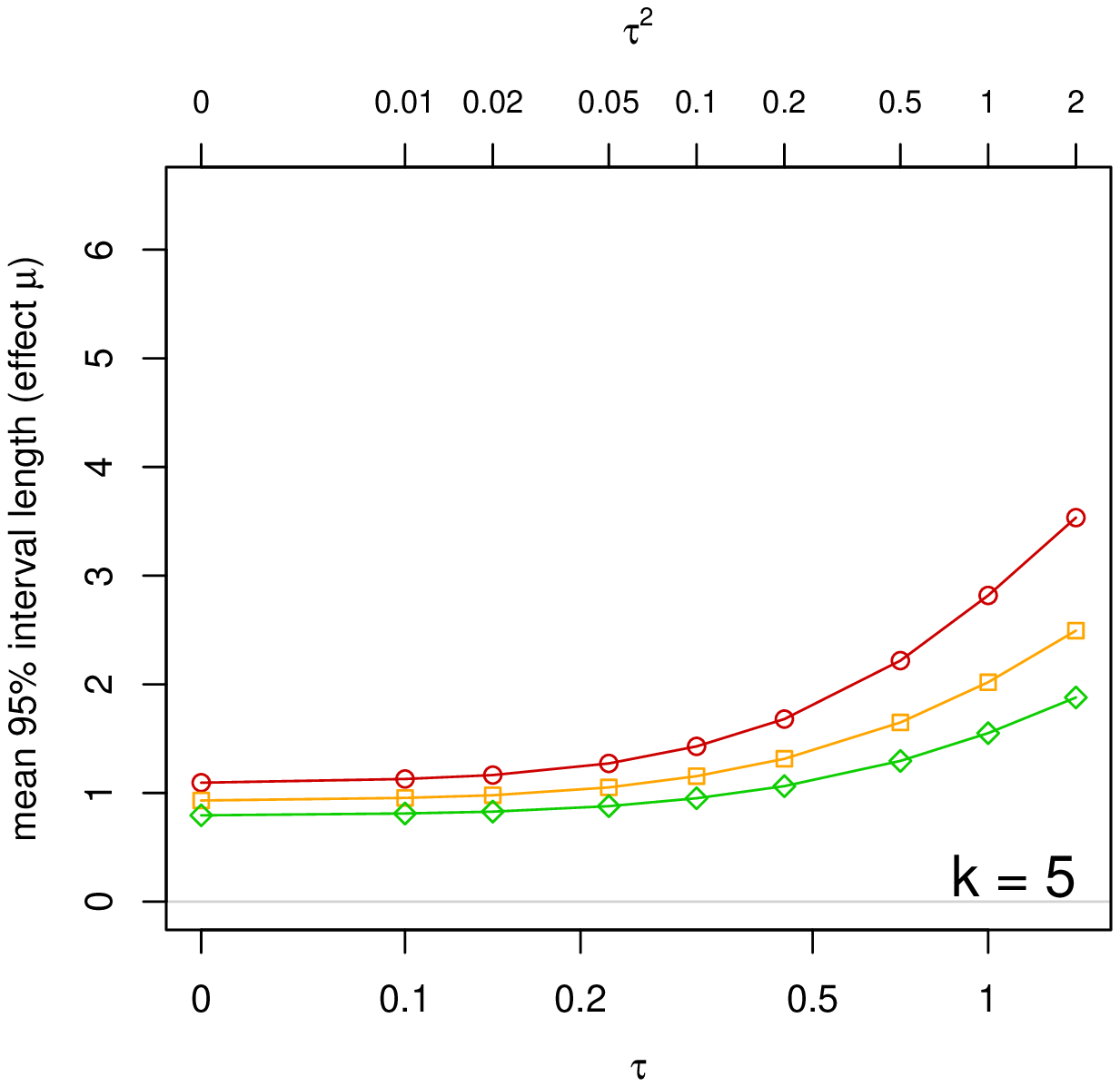} 
\includegraphics[width=0.3\linewidth]{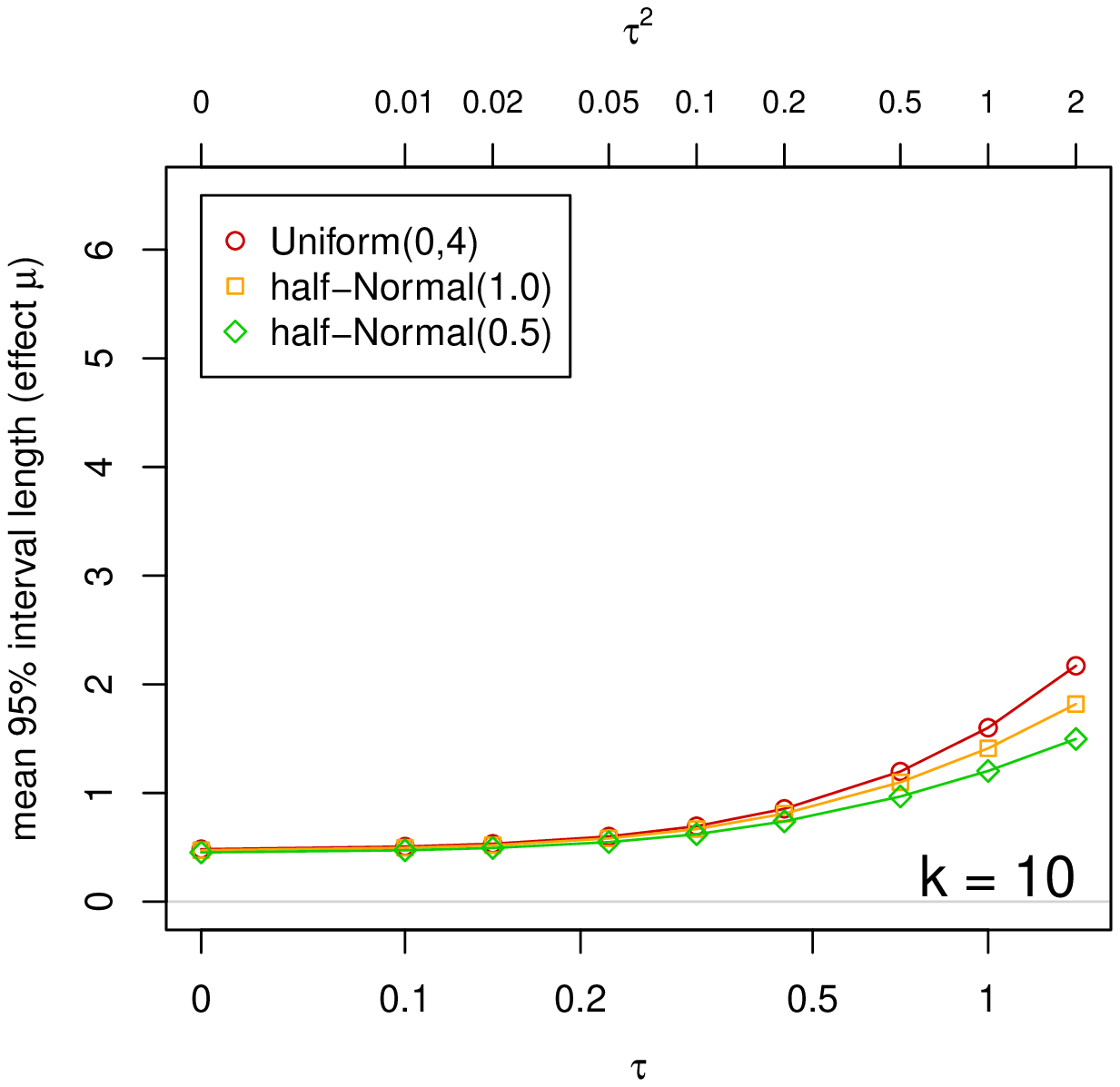} 
\caption{Mean lengths of the credibility / confidence intervals for the overall treatment effect~$\mu$ depending on the between-study heterogeneity $\tau$ and several numbers $k$ of studies included in the meta-analyses using various estimators.} \label{fig:cilength}
\end{center}
\end{figure}

Figure~\ref{fig:cilength} shows the simulated mean lengths of several credibility and confidence intervals for the treatment effect~$\mu$ depending on the between-trial heterogeneity~$\tau^2$ and the number of included studies~$k$. By construction the confidence intervals based on normal approximation are shorter than the Knapp-Hartung intervals. The Bayesian intervals tend to be shorter than the Knapp-Hartung intervals, also in situations with similar or even larger coverage probability.

In conclusion, the method based on normal quantiles cannot be recommended for use with small numbers of studies, since the observed coverage was often well below the nominal level. The Knapp-Hartung approach leads to coverage at or above the nominal level, but in cases with very few studies results in extremely long confidence intervals. Bayesian credibility intervals with appropriate choice of prior performed well and can be recommended for application in rare diseases. However, with very small number of studies included the results are sensitive to the prior specification.

\section{The motivating case study revisited} \label{sec:example_rev}

In this section we are returning to the case study by \citet{CrinsEtAl2014} introduced in Section~\ref{sec:example}. Crins et al.\ investigated immunosuppressive therapies in paediatric liver transplantation with regard to acute and steroid-resistant rejections. In the first part of this section we will apply the methods introduced in Section~\ref{sec:meth} to these data and discuss the results. In the second part we will investigate the long-run behaviour of the different meta-analysis procedures based on scenarios motivated by the liver transplant example simulating binomial data rather than normal test statistics and their standard errors.

\begin{figure}[ht]
\begin{center}
\includegraphics[width=0.48\linewidth]{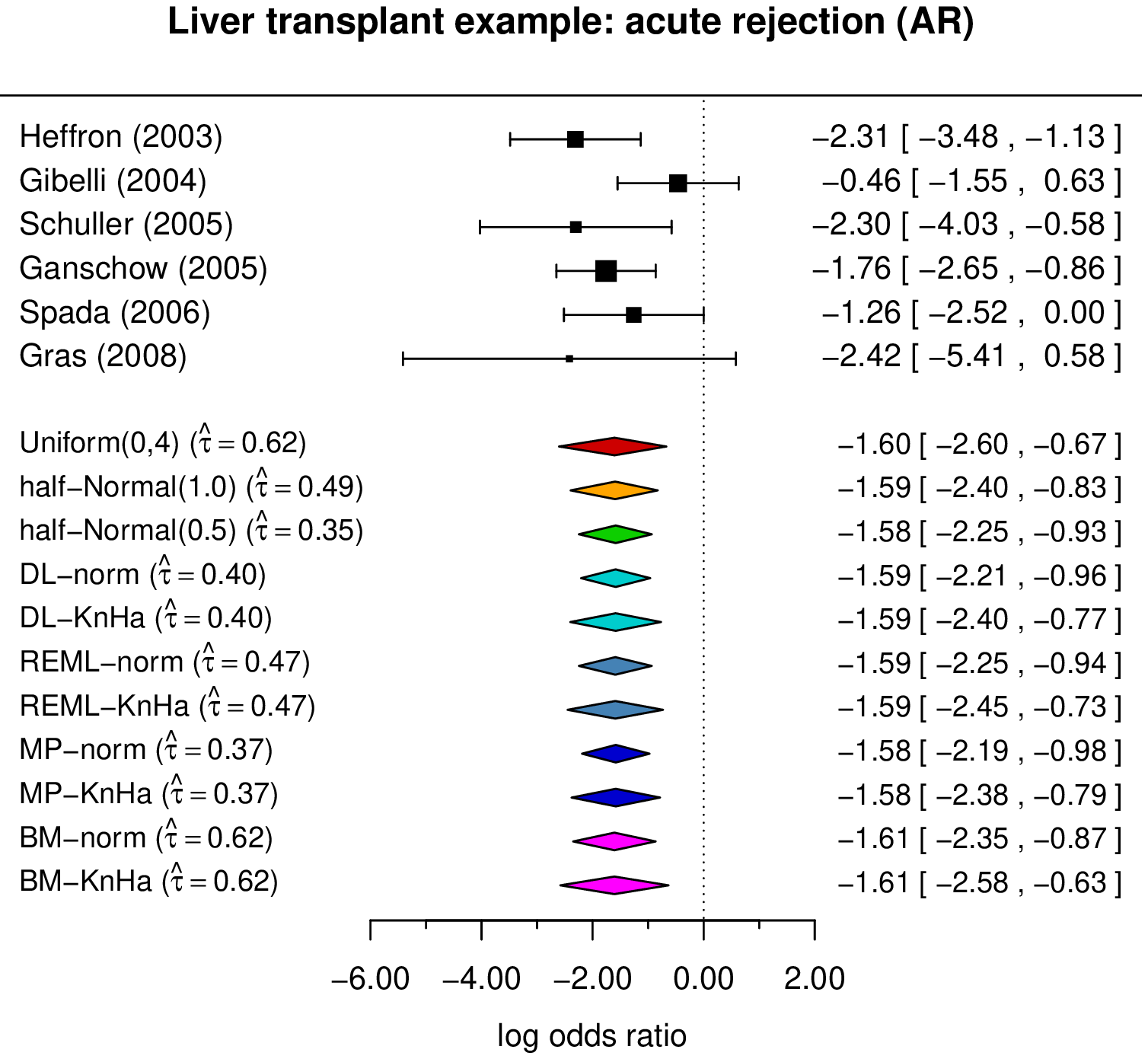}
\includegraphics[width=0.48\linewidth]{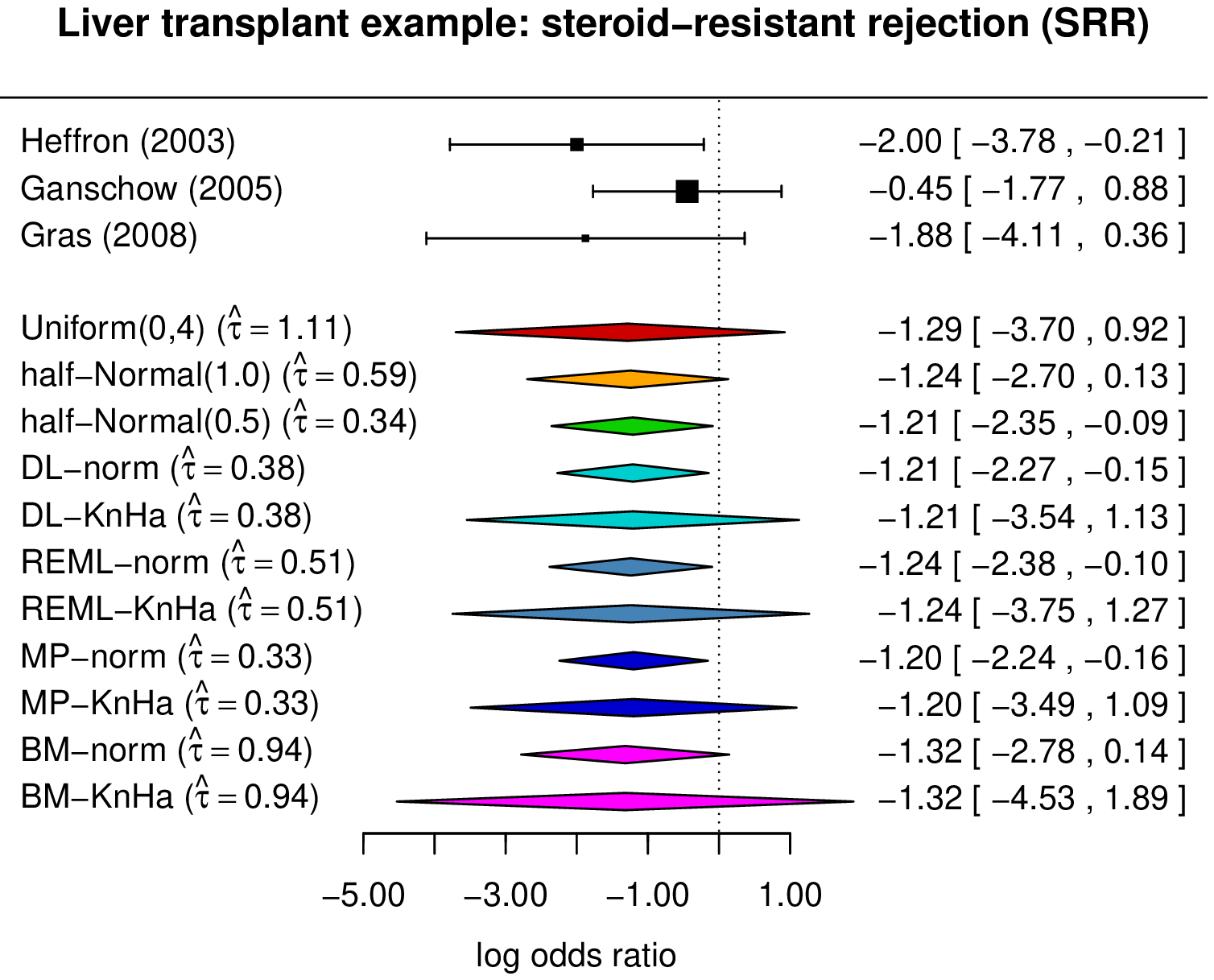}
\caption{Paediatric transplant example: Treatment effect estimates on the log odds ratio scale with 95\% confidence intervals for the individual studies and meta-analyses as well as estimates $\hat \tau$ of the between-trial heterogeneity. For the Bayes methods the posterior medians are given.}\label{fig:example_forest}
\end{center}
\end{figure}

Figure~\ref{fig:example_forest} shows the results of applying the various methods introduced in Section~\ref{sec:meth} to the paediatric liver transplant data presented in Section~\ref{sec:example}. In the figure the treatment effect estimates are shown on the log odds ratio scale with 95\% CI for the individual studies as well as for the meta-analyses. The estimates $\hat \tau$ of the between-trial heterogeneity are also included in the figure. For the Bayes methods the posterior medians are given. Whereas in the case of acute rejections the results are fairly similar across the various methods with statistically significant combined treatment effects as reported in \citet{CrinsEtAl2014}, there are some marked differences between the methods in the case of steroid-resistant rejections. On the whole the methods based on normal approximation lead to the shortest confidence intervals which are indicating statistically significant effects with the exception of the \textsl{BM} method. The Knapp-Hartung approach leads to intervals including 0 independent of the heterogeneity estimation method and are among the longest intervals obtained. With only three studies included in the SRR meta-analyses, the differences between the Knapp-Hartung and the normal approaches are more pronounced than in the AR example. In this example the Bayes methods provide a middle ground with credibility intervals which are on the whole shorter than those obtained by the Knapp-Hartung method and upper credibility limits close to 0 with the only exception being the estimate based on the conservative uniform prior. However, due to the very small number of studies the Bayes estimators are more sensitive to the specification of prior than in the AR example with six trials.

In the AR example the estimates of the between-trial heterogeneity $\tau$ center around 0.5 and vary from 0.37 with the \textsl{MP} estimator to 0.62 for the \textsl{BM} estimator and the Bayes estimator with uniform prior on the interval from 0 to 4. In the SRR data the observed heterogeneity is somewhat higher in comparison with estimates ranging from 0.33 for the \textsl{MP} estimator to 1.11 for the Bayes estimator with uniform prior. All these estimates come with a considerable degree of uncertainty reflected in the length of the confidence / credibility intervals. For the acute rejections the 95\% credibility intervals of $\tau$ are 0 to 1.848 with uniform prior, to 1.260 with half-normal prior with scale 1, and to 0.862 for half-normal prior with scale 0.5 whereas the 95\% confidence interval using the Q-profile method extends from 0 to 1.726. For the steroid-resistant rejections the 95\% credibility intervals range from 0 to 3.368, 1.652, and 0.941 for the uniform and the two half-normal priors with scales 1 and 0.5, respectively. The 95\% Q-profile interval even extends from 0 to 5.365. As can be seen from this example, the credibility intervals depend very much on the choice of prior, in particular with very small numbers of studies. The frequentist Q-profile method gives confidence intervals as wide or wider than the credibility intervals based on the conservative uniform prior.

We exemplarily investigated the long-run behaviour of the different meta-analysis procedures based on scenarios motivated by the liver transplant example. For this purpose we simulated data similar to the meta-analyses of acute rejections and steroid-resistant rejections. The former included six studies, the latter included only three. In each simulation iteration, first true group-specific odds were generated as normally distributed on the log-odds scale; for AR with means~$\mu_\mathrm{control}=0.0$ and~$\mu_\mathrm{treatment}=-1.5$ and for SR with $\mu_\mathrm{control}=-2.0$ and $\mu_\mathrm{treatment}=-3.0$. Variances were equal to~$1$, and the log-odds were positively correlated with correlation coefficient $0.875$ and $0.719$ for AR and SRR, respectively. The resulting between-trial heterogeneity in terms of log-odds ratios is $\tau=0.5$ for ARR and $\tau=0.75$ for SRR, which are substantial and substantial to large levels of hetereogeneity according to Table~\ref{tab:tau}. Resulting log-odds were then used to generate study results (contingency tables) based on a binomial distribution and patient numbers as given in the examples. The corresponding effect size estimates and standard errors were computed using the {\tt escalc()} function from the {\tt metafor} package \citep{Viechtbauer2010}, which were then analyzed using the different methods. The results of these simulations are summarized in Table~\ref{tab:crins} where the coverage probabilities and mean lengths of the 95\% confidence intervals for the treatment effect~$\mu$ are given. As with the simulation study presented in Section~\ref{sec:sims}, 10,000 trials were simulated for each scenario. 

Looking at Table~\ref{tab:crins} it is apparent that the confidence intervals based on the normal quantiles have coverages below the nominal level of 0.95 with the exception of the interval using the \textsl{BM} estimator of the between-trial heterogeneity $\tau$, which is slightly conservative in both scenarios. The Knapp-Hartung confidence intervals are in comparison much longer and fairly conservative. In the SRR scenario this is more extreme due to the very small number of trials. Although the true between-trial heterogeneity is substantial ($\tau=0.5$) in the AR scenario and substantial to large ($\tau=0.75$) in the SRR scenario the proportion of estimating 0 for $\tau$ is relatively large, explaining partly the disappointing performance of some frequentist procedures. In the ARR scenario the proportions are 0.333, 0.255, and 0.292 with the \textsl{DL}, \textsl{REML} and \textsl{MP} estimator, respectively. In the SRR setting the proportions are 0.523, 0.447, and 0.487. The interval with the coverage probabilities closest to the nominal level of 0.95 out of all interval estimators investigated is the Bayes credibility interal with half normal prior and scale 0.5. The more conservative choices of prior distributions of $\tau$, i.e. uniform on the interval from 0 to 4 and half-normal with scale 1, lead to longer and more conservative credibility intervals. Again the dependence of the coverage probabilities on the the prior selected is more pronounced in the SRR scenario with fewer studies than in the AR scenario.

\begin{table}[ht]
\centering
\caption{Comparison of different methods using simulations of binary data imitating the example settings. Coverage probabilities (mean lengths) of 95\% confidence intervals for the treatment effect $\mu$ are given.} \label{tab:crins}
\begin{tabular}{lcc}
  \hline
\hline
 & AR & SRR \\ 
  \hline
  DL-norm & 93.1 (1.28) & 91.7 (2.51) \\ 
  REML-norm & 92.8 (1.27) & 91.7 (2.50) \\ 
  EB-norm & 93.2 (1.29) & 91.7 (2.51) \\ 
  BM-norm & 96.7 (1.46) & 97.9 (3.24) \\ 
   &&\\[-1ex]
  DL-KnHa & 98.0 (1.71) & 99.9 (5.58) \\ 
  REML-KnHa & 98.1 (1.71) & 100.0 (5.58) \\ 
  EB-KnHa & 98.0 (1.70) & 99.9 (5.52) \\ 
  BM-KnHa & 99.4 (1.92) & 100.0 (7.12) \\ 
   &&\\[-1ex]
  Uniform(0,4) & 99.0 (1.91) & 100.0 (4.84) \\ 
  half-Normal(1.0) & 97.8 (1.55) & 98.0 (3.00) \\ 
  half-Normal(0.5) & 95.5 (1.30) & 94.2 (2.38) \\ 
   \hline
\hline
\end{tabular}
\end{table}

The results of this small simulation generating data from binomial distributions tie in nicely with the extensive simulation study presented in Section~\ref{sec:sims} where effect estimates and their standard errors were sampled directly. In comparing back, one should note that the magnitude of the heterogeneity $\tau$ may not be directly comparable to the values used in the simulations in Section~\ref{sec:sims}. In the simulations, the studies' standard errors~$\sigma_i$ were drawn from a distribution ranging from 0.095 to 0.775 (with median 0.35), while in the paediatric transplantation example, the standard errors ranged from 0.056 to 1.529 for AR and from 0.676 to 1.142 for SRR, so that the same absolute heterogeneity~$\tau$ corresponds to a smaller \textsl{relative} amount of heterogeneity (e.g.~$I^2$) in the transplantation example than in the previous simulations.

\section{Discussion}
In rare diseases and small populations data and biomaterial of an individual patient are even more precious due to their rarity. In these circumstances the synthesis of the evidence available is an important step in the development of new treatments. Meta-analyses in rare diseases and small populations face particular problems which can be characterized by a small number of studies included in the meta-analysis, relatively small study sizes, and heterogeneity between study results which might be due to a variety of study designs employed. In this paper we reviewed a number of frequentist and Bayesian procedures and investigated their suitability for meta-analyses in rare diseases and small populations. 

In summary, we found that frequentist estimators of the between-trial heterogeneity often fail to pick up the variation present in the data, in particular when the number of trials is very small (e.g.\ $k=3$), resulting in a relatively high proportion of estimates being 0. The observed downwards bias is in agreement with previous findings (see e.g.\ \citet{ChungEtAl2013a,BoehningEtAl2002}). Furthermore, the commonly applied approach of constructing confidence intervals using normal quantiles fails to reflect the uncertainty in the estimation of the between-trial heterogeneity. This has been recognized before and methods based on $t$-distributions have been proposed to tackle the problem \citep{FollmannProschan1999,HartungKnapp2001a,HartungKnapp2001b,KnappHartung2003,HigginsThompsonSpiegelhalter2009}. Here we included the method by Knapp-Hartung in our investigations. Their approach yields more favourable results in terms of coverage probabilities of the confidence intervals than the standard approach based on normal quantiles \citep{HartungKnapp2001a,HartungKnapp2001b}. As we have seen in the simulations presented here, the Knapp-Hartung approach controls the coverage probability even in settings with very few small studies, which was also demonstrated in a very recent commentary for the case of only two studies ($k=2$) \citep{GonnnermannEtAl2015}. As we have seen here, however, these intervals are often very long and fairly conservative, which again is in agreement with previous findings \citep{GonnnermannEtAl2015,HarbordHiggins2008}.

The Bayesian credibility intervals based on uniform and half-normal priors also exhibited coverage probabilities in excess of the nominal level for a range of scenarios considered. However, they tended to be shorter than those obtained by the Knapp-Hartung method while producing coverage probabilities either similar to the Knapp-Hartung method or closer to the nominal level. With very few studies the performance of the Bayesian credibility intervals is of course sensitive to the specification of the prior for the between-trial heterogeneity \citep{LambertEtAl2005}. On the whole the proposed half-normal priors with scales 0.5 or 1 performed well across the scenarios considered, which we believe reflect typical situations encountered in rare diseases as evidenced by our example in immunosuppression after liver transplantation in children. Therefore, we would recommend this choice of prior for applications in rare diseases and small populations. Furthermore, these priors are also supported by empirical evidence reported in \citet{TurnerEtAl2015}. With larger numbers of studies of course the prior specification becomes less important.   

Frequentist and Bayesian appproaches to interval estimation in meta-analyses have been compared elsewhere \citep{ChungEtAl2013a}. However, our comparisons were not only different by focussing on scenarios typical for rare diseases and small populations, but also in the range of scenarios considered and number of simulation replications performed. We were able to run this relatively extensive simulation study including computationally intensive fully Bayesian approaches because of an advance in their implementation resulting in the \textsf{R} package \texttt{bayesmeta} moving away, as very recently suggested in \citet{TurnerEtAl2015}, from standard Markov chain Monte Carlo techniques.

Our investigations were limited to a small number of estimators of the between-trial heterogeneity $\tau$. These included standard estimators commonly applied in meta-analyses such as the DerSimonian-Laird estimator, but also the recently proposed Bayes-Modal estimator by \citep{ChungEtAl2013a}. However, a number of alternative estimators of $\tau$ have been proposed which have been compared elsewhere \citep{VeronikiEtAl2015}. Furthermore, we focused here on meta-analyses including a small number of studies ($k\le 10$) which was motivated by rare diseases and small populations. However, this is also otherwise not uncommon as a review by Turner et al.\ of the Cochrane Database shows \citep{TurnerEtAl2012}. We also restricted simulations to normally distributed outcomes, without considering the effects of transforming other types of outcomes to a continuous scale. Such effects become especially relevant when approximate normality breaks down, as for example when dealing with data on rare events \citep{BradburnEtAl2007}.

\section*{Acknowledgements}
This research has received funding from the EU's 7th Framework Programme for research, technological development and demonstration under grant agreement number FP HEALTH 2013-602144 with project title (acronym) ``Innovative methodology for small populations research'' (InSPiRe).

{
  \small
  \bibliographystyle{plainnat}
  \bibliography{literature}
}

\end{document}